\documentclass[pdftex, a4paper, twoside]{article}
\usepackage{amsmath,amssymb,bm}
\usepackage{caption}
\usepackage{color}
\usepackage{wrapfig}
\usepackage{graphicx}
\usepackage{fancyhdr}
\usepackage{a4wide}
\usepackage{ae,aecompl,aeguill}
\usepackage{microtype}
\usepackage{hyperref}

\definecolor{DarkBlue}{rgb}{0,0.08,0.45}
\hypersetup{%
  a4paper,
  pdfauthor={Carsten Rostgaard},
  pdftitle={The Projector Augmented-wave Method},
  pdfsubject={The Projector Augmented-wave Method},
  pdfstartpage=1,
  colorlinks=true,
  pdfstartview=Fit,
  linkcolor=DarkBlue,
  citecolor=DarkBlue,
  urlcolor=DarkBlue,
  bookmarksnumbered=true,
  hypertexnames=true,
  plainpages=false,
  linkbordercolor={0 0 1}}

\newcommand{\gpaw}{\textsc{gpaw}}

\newcommand{\f}[1]{\mathbf{#1}}

\newcommand{\s}[1]{\tilde{#1}}
\newcommand{\ws}[1]{\widetilde{#1}}
\newcommand{\h}[1]{\hat{#1}}
\newcommand{\wh}[1]{\widehat{#1}}
\newcommand{\ext}{\text{ext}}
\newcommand{\br}{\mathbf{r}}
\newcommand{\rr}{|\mathbf{r} - \mathbf{r}'|}

\newcommand{\bR}{\mathbf{R}}
\newcommand{\T}{\hat{\mathcal{T}}}
\newcommand{\Z}{\mathcal{Z}}
\newcommand{\Ham}{\widehat{H}}
\newcommand{\bra}[1]{\langle #1 |}
\newcommand{\ket}[1]{| #1 \rangle}
\newcommand{\braket}[2]{\langle #1 | #2 \rangle}
\newcommand{\psit}{\tilde{\psi}}

\newcommand{\pt}{\tilde{p}}

\DeclareMathOperator{\tr}{Tr}

\title{\textbf{The Projector Augmented-wave Method}}
\author{Carsten Rostgaard}
\date{October 10, 2009}

\begin{document}
\thispagestyle{empty}
\maketitle
\begin{abstract}
  The purpose of this text is to give a self-contained description of
  the basic theory of the projector augmented-wave (PAW) method, as
  well as most of the details required to make the method work in
  practice. These two topics are covered in the first two sections,
  while the last is dedicated to examples of how to apply the PAW
  transformation when extracting non-standard quantities from a
  density-functional theory (DFT) calculation.

  The formalism is based on Bl{\"o}chl's original formulation of PAW
  \cite{Blochl1994}, and the notation and extensions follow those used
  and implemented in the \gpaw\cite{gpaw} code.
\end{abstract}
\tableofcontents
\clearpage

\section{Formalism}\label{sec: paw}
By the requirement of orthogonality, DFT wave functions have very
sharp features close to the nuclei, as all the states are non-zero in
this region. Further out only the valence states are non-zero,
resulting in much smoother wave functions in this interstitial region.
The oscillatory behavior in the core regions, requires a very large
set of plane waves, or equivalently a very fine grid, to be described
correctly. One way of solving this problem is the use of
pseudopotentials in which the collective system of nuclei and core
electrons are described by an effective, much smoother, potential.
The KS equations are then solved for the valence electrons only.  The
pseudopotentials are constructed such that the correct scattering
potential is obtained beyond a certain radius from the core. This
method reduces the number of wave functions to be calculated, since
the pseudo potentials only have to be calculated and tabulated once
for each type of atom, so that only calculations on the valence states
are needed. It justifies the neglect of relativistic effects in the KS
equations, since the valence electrons are non-relativistic (the
pseudopotentials describing core states are of course constructed with
full consideration of relativistic effects). The technique also
removes the unwanted singular behavior of the ionic potential at the
lattice points.
\par The drawback of the method is that all information on the full
wave function close to the nuclei is lost. This can influence the
calculation of certain properties, such as hyperfine parameters, and
electric field gradients. Another problem is that one has no before
hand knowledge of when the approximation yields reliable results.
\par A different approach is the augmented-plane-wave method
(APW), in which space is divided into atom-centered augmentation
spheres inside which the wave functions are taken as some
atom-like partial waves, and a bonding region outside the spheres,
where some envelope functions are defined. The partial waves and
envelope functions are then matched at the boundaries of the
spheres.
\par A more general approach is the projector augmented wave method
(PAW) presented here, which offers APW as a special
case\cite{Blochl2003}, and the pseudopotential method as a well
defined approximation\cite{Kresse1999}. The PAW method was first
proposed by Blöchl in 1994\cite{Blochl1994}.

\subsection{The Transformation Operator}\label{sec: transformation operator}
The features of the wave functions, are very different in different
regions of space. In the bonding region it is smooth, but near the
nuclei it displays rapid oscillations, which are very demanding on the
numerical representation of the wave functions.  To address this
problem, we seek a linear transformation $\T$ which takes us from an
auxiliary smooth wave function $\ket{\s{\psi}_n}$ to the true all
electron Kohn-Sham single particle wave function $\ket{\psi_n}$
\begin{equation}
\ket{\psi_n}=\T\ket{\s{\psi}_n},
\end{equation}
where $n$ is the quantum state label, containing a $\f{k}$ index, a
band index, and a spin index.
\par This transformation yields the transformed KS equations
\begin{equation}\label{eq: paw ks equations}
\T^\dagger \Ham \T \ket{\s{\psi}_n}=\epsilon_n\T^\dagger\T\ket{\s{\psi}_n},
\end{equation}
which needs to be solved instead of the usual KS equation. Now we need
to define $\T$ in a suitable way, such that the auxiliary wave
functions obtained from solving \eqref{eq: paw ks equations} becomes
smooth.
\par Since the true wave functions are already smooth at a certain
minimum distance from the core, $\T$ should only modify the wave
function close to the nuclei. We thus define
\begin{equation}
\T = 1 + \sum_a \T^a,
\end{equation}
where $a$ is an atom index, and the atom-centered transformation,
$\T^a$, has no effect outside a certain atom-specific augmentation
region $|\f{r}-\f{R}^a|<r_c^a$. The cut-off radii, $r_c^a$ should
be chosen such that there is no overlap of the augmentation
spheres.
\par Inside the augmentation spheres, we expand the true wave function in the partial waves
$\phi_i^a$, and for each of these partial waves, we define a
corresponding auxiliary smooth partial wave $\s{\phi}_i^a$, and
require that
\begin{equation}\label{eq: ta}
\ket{\phi_i^a}
=(1+\T^a)\ket{\s{\phi}_i^a}\quad\Leftrightarrow\quad
\T^a\ket{\s{\phi}_i^a} =\ket{\phi_i^a} - \ket{\s{\phi}_i^a}
\end{equation}
for all $i,a$. This completely defines $\T$, given $\phi$ and $\s{\phi}$.
\par Since $\T^a$
should do nothing outside the augmentation sphere, we see from
\eqref{eq: ta} that we must require the partial wave and its
smooth counterpart to be identical outside the augmentation sphere
\begin{equation*}
\forall a, \quad \phi_i^a(\f{r})=\s{\phi}_i^a(\f{r})\text{, for } r>r_c^a
\end{equation*}
where $\phi_i^a(\f{r})=\braket{\f{r}}{\phi_i^a}$ and likewise for
$\s{\phi}_i^a$.
\par If the smooth partial waves form a complete set inside the
augmentation sphere, we can formally expand the smooth all electron
wave functions as
\begin{equation}\label{eq: smooth psi expansion}
\ket{\s{\psi}_n} = \sum_i P_{ni}^a \ket{\s{\phi}_i^a} \text{, for } |\f{r}-\f{R}^a|<r_c^a
\end{equation}
where $P_{ni}^a$ are some, for now, undetermined expansion
coefficients.
\par Since $\ket{\phi_i^a} =
\T\ket{\s{\phi}_i^a}$ we see that the expansion
\begin{equation}\label{eq: psi expansion}
\ket{\psi_n} = \T\ket{\s{\psi}_n} = \sum_i P_{ni}^a \ket{\phi_i^a} \text{, for } |\f{r}-\f{R}^a|<r_c^a
\end{equation}
has identical expansion coefficients, $P_{ni}^a$.
\par As we require $\T$ to be linear, the coefficients $P_{ni}^a$ must be
linear functionals of $\ket{\s{\psi}_n}$, i.e.
\begin{equation}\label{eq: P expansion coeff}
P_{ni}^a = \braket{\s{p}_i^a}{\s{\psi}_n} = \int d\f{r}
\s{p}_i^{a*}(\f{r}-\f{R}^a)\s{\psi}_n(\f{r}),
\end{equation}
where $\ket{\s{p}_i^a}$ are some fixed functions termed smooth
projector functions.
\par As there is no overlap between the augmentation spheres, we
expect the one center expansion of the smooth all electron wave
function, $\ket{\s{\psi}^a_n} = \sum_i
\ket{\s{\phi}_i^a}\braket{\s{p}_i^a}{\s{\psi}_n}$ to reduce to
$\ket{\s{\psi}_n}$ itself inside the augmentation sphere defined by
$a$. Thus, the smooth projector functions must satisfy
\begin{equation}\label{eq: phi p completeness}
\sum_i \ket{\s{\phi}_i^a}\bra{\s{p}_i^a} = 1 
\end{equation}
inside each augmentation sphere. This also implies that
\begin{equation}\label{eq: phi p orthogonal}
\braket{\s{p}_{i_1}^a}{\s{\phi}_{i_2}^a}=\delta_{i_1,i_2} \text{~, for } |\f{r}-\f{R}^a|<r_c^a
\end{equation}
i.e. the projector functions should be orthonormal to the smooth
partial waves inside the augmentation sphere. There are no
restrictions on $\s{p}_i^a$ outside the augmentation spheres, so
for convenience we might as well define them as local functions,
i.e. $\s{p}_i^a(\f{r})=0$ for $r>r_c^a$.
\par Note that the completeness relation \eqref{eq: phi p
completeness} is equivalent to the requirement that $\s{p}_i^a$
should produce the correct expansion coefficients of \eqref{eq:
smooth psi expansion}-\eqref{eq: psi expansion}, while \eqref{eq:
phi p orthogonal} is merely an implication of this restriction.
Translating \eqref{eq: phi p completeness} to an explicit
restriction on the projector functions is a rather involved
procedure, but according to Blöchl, \cite{Blochl1994}, the most
general form of the projector functions is:
\begin{equation}\label{eq: projector general}
\bra{\s{p}_i^a} = \sum_j
(\{\braket{f_k^a}{\s{\phi}_l^a}\})^{-1}_{ij}\bra{f_j^a},
\end{equation}
where $\ket{f_j^a}$ is any set of linearly independent functions.
The projector functions will be localized if the functions
$\ket{f_j^a}$ are localized.
\par Using the completeness relation \eqref{eq: phi p
completeness}, we see that
\begin{equation*}
\T^a =\sum_i \T^a\ket{\s{\phi}_i^a}\bra{\s{p}_i^a} = \sum_i
\big(\ket{\phi_i^a} - \ket{\s{\phi}_i^a}\big) \bra{\s{p}_i^a},
\end{equation*}
where the first equality is true in  all of space, since
\eqref{eq: phi p completeness} holds inside the augmentation
spheres and outside $\T^a$ is zero, so anything can be multiplied
with it. The second equality is due to \eqref{eq: ta} (remember
that $\ket{\phi_i^a} - \ket{\s{\phi}_i^a}=0$ outside the
augmentation sphere). Thus we conclude that
\begin{equation}\label{eq: T operator}
\T =1+ \sum_a\sum_i \big(\ket{\phi_i^a} - \ket{\s{\phi}_i^a}\big)
\bra{\s{p}_i^a}.
\end{equation}
\par To summarize, we obtain the all electron KS wave function
$\psi_n(\f{r})=\braket{\f{r}}{\psi_n}$ from the transformation
\begin{equation}\label{eq: psi transform in r}
\psi_n(\f{r}) = \s{\psi}_n(\f{r})+\sum_a\sum_i \big( \phi_i^a(\f{r})
- \s{\phi}_i^a(\f{r}) \big)\braket{\s{p}_i^a}{\s{\psi}_n},
\end{equation}
where the smooth (and thereby numerically convenient) auxiliary wave
function $\s{\psi}_n(\f{r})$ is obtained by solving the eigenvalue
equation \eqref{eq: paw ks equations}.
\par The transformation \eqref{eq: psi transform in r} is expressed in
terms of the three components: a) the partial waves $\phi_i^a(\f{r})$,
b) the smooth partial waves $\s{\phi}_i^a(\f{r})$, and c) the smooth
projector functions $\s{p}_i^a(\f{r})$.
\par The restriction on the choice of these sets of functions are: a)
Since the partial- and smooth partial wave functions are used to
expand the all electron wave functions, i.e. are used as atom-specific
basis sets, these must be complete (inside the augmentation spheres).
b) the smooth projector functions must satisfy \eqref{eq: phi p
  completeness}, i.e. be constructed according to \eqref{eq: projector
  general}.  All remaining degrees of freedom are used to make the
expansions converge as fast as possible, and to make the functions
termed `smooth', as smooth as possible. For a specific choice of these
sets of functions, see section \ref{sec: partial wave basis}. As the
partial- and smooth partial waves are merely used as basis sets they
can be chosen as real functions (any imaginary parts of the functions
they expand, are then introduced through the complex expansion
coefficients $P_{ni}^a$).  In the remainder of this document $\phi$
and $\s{\phi}$ will be assumed real.
\par Note that the sets of functions needed to define the transformation are
system independent, and as such they can conveniently be
pre-calculated and tabulated for each element of the periodic
table.
\par For future convenience, we also define the one center expansions
\begin{subequations}
\begin{align}
\psi_{n}^a(\f{r}) &= \sum_i \phi_i^a(\f{r})\braket{\s{p}_i^a}{\s{\psi}_n},\\
\s{\psi}_{n}^a(\f{r}) &= \sum_i
\s{\phi}_i^a(\f{r})\braket{\s{p}_i^a}{\s{\psi}_n}.
\end{align}
\end{subequations}
In terms of these, the all electron KS wave function is
\begin{equation}
\psi_n(\f{r})=\s{\psi}_n(\f{r})+\sum_{a} \big(
\psi_{n}^{a}(\f{r}-\f{R}^a) - \s{\psi}_{n}^a(\f{r}-\f{R}^a) \big).
\end{equation}

\par So what have we achieved by this transformation? The trouble of
the original KS wave functions, was that they displayed rapid
oscillations in some parts of space, and smooth behavior in other
parts of space. By the decomposition \eqref{eq: psi transform in r} we
have separated the original wave functions into auxiliary wave
functions which are smooth everywhere and a contribution which
contains rapid oscillations, but only contributes in certain, small,
areas of space. This decomposition is shown on the front page for the
hydrogen molecule. Having separated the different types of waves,
these can be treated individually. The localized atom centered parts,
are indicated by a superscript $a$, and can efficiently be represented
on atom centered radial grids. Smooth functions are indicated by a
tilde \~{}. The delocalized parts (no superscript $a$) are all smooth,
and can thus be represented on coarse Fourier- or real space grids.

\subsection{The Frozen Core Approximation}\label{sec: frozen core}
In the frozen core approximation, it is assumed that the core states
are naturally localized within the augmentation spheres, and that the
core states of the isolated atoms are not changed by the formation of
molecules or solids. Thus the core KS states are identical to the
atomic core states
\begin{wrapfigure}[17]{r}[35pt]{0pt}
\includegraphics[scale=0.56]{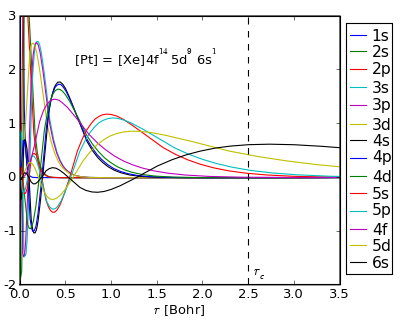}%
\caption{The core states of Platinum}\label{fig: frozen core}%
\end{wrapfigure}
\begin{equation*}
\ket{\psi^c_n} = \ket{\phi_\alpha^{a,\text{core}}},
\end{equation*}
where the index $n$ on the left hand site refers to both a specific
atom, $a$, and an atomic state, $\alpha$.
\par In this approximation only valence states are included in the
expansions of $\ket{\psi_n}$, \eqref{eq: psi expansion}, and
$\ket{\s{\psi}_n}$, \eqref{eq: smooth psi expansion}.
\par Figure \ref{fig: frozen core}, shows the atomic states of
Platinum in its ground state, obtained with an atomic DFT program at
an LDA level, using spherical averaging, i.e. a spin-compensated
calculation, assuming the degenerate occupation 9/10 of all 5d states,
and both of the 6s states half filled. It is seen that at the typical
length of atomic interaction (the indicated cut-off $r_c=2.5$ Bohr is
approximately half the inter-atomic distance in bulk Pt), only the 5d
and 6s states are non-zero.

\subsection{Expectation Values}\label{sec: expectation values}

The expectation value of an operator $\wh{O}$ is, within the frozen
core approximation, given by
\begin{equation}
  \langle \wh{O} \rangle = \sum_n^\text{val} f_n \bra{\psi_n}\wh{O}\ket{\psi_n} + \sum_a \sum_\alpha^\text{core} \bra{\phi_\alpha^{a,\text{core}}}\wh{O}\ket{\phi_\alpha^{a,\text{core}}}.
\end{equation}
Using that $\bra{\psi_n}\wh{O}\ket{\psi_n} =
\bra{\s{\psi}_n}\T^\dagger \wh{O}\T\ket{\s{\psi}_n}$, and skipping the
state index for notational convenience, we see that
\begin{equation}
  \begin{split}
    \bra{\psi}\wh{O}\ket{\psi} &= \bra{\s{\psi} + \sum_a(\psi^a - \s{\psi}^a)} \wh{O} \ket {\s{\psi} + \sum_a(\psi^a - \s{\psi}^a)}\\
&= \bra{\s{\psi}}\wh{O}\ket{\s{\psi}} + \sum_{aa'} \bra{\psi^a - \s{\psi}^a}\wh{O}\ket{\psi^{a'} - \s{\psi}^{a'}} + \sum_a \left(\bra{\s{\psi}}\wh{O}\ket{\psi^a - \s{\psi}^a} + \bra{\psi^a - \s{\psi}^a}\wh{O}\ket{\s{\psi}}\right)\\
&= \bra{\s{\psi}}\wh{O}\ket{\s{\psi}} + \sum_{a}\left( \bra{\psi^a}\wh{O}\ket{\psi^a} - \bra{\s{\psi}^a}\wh{O}\ket{\s{\psi}^a} \right)\\
& \hspace{48pt}+ \sum_a \left( \bra{\psi^a - \s{\psi}^a}\wh{O}\ket{\s{\psi} - \s{\psi}^a} + \bra{\s{\psi} - \s{\psi}^a}\wh{O}\ket{\psi^a - \s{\psi}^a} \right)\\
& \hspace{48pt}+ \sum_{a\neq a'} \bra{\psi^a - \s{\psi}^a}\wh{O}\ket{\psi^{a'} - \s{\psi}^{a'}}.
  \end{split}
\end{equation}
For local operators\footnote{Local operator $\wh{O}$: An operator
  which does not correlate separate parts of space, i.e.
  $\bra{\br}\wh{O}\ket{\br'} = 0$ if $\br\neq \br'$.} the last two
lines does not contribute. The first line, because $\ket{\psi^a -
  \s{\psi}^a}$ is only non-zero inside the spheres, while
$\ket{\s{\psi} - \s{\psi}^a}$ is only non-zero outside the spheres.
The second line simply because $\ket{\psi^a - \s{\psi}^a}$ is zero
outside the spheres, so two such states centered on different nuclei
have no overlap (provided that the augmentation spheres do not
overlap).
\par Reintroducing the partial waves in the one-center expansions, we see that
\begin{equation}
\sum_n^\text{val} f_n \bra{\psi_{n}^{a}}\wh{O}\ket{\psi_{n}^{a}} = \sum_n^\text{val} f_n \sum_{i_1i_2} \bra{\phi_{i_1}^{a}P^a_{ni_1}} \wh{O} \ket{\phi_{i_2}^{a} P_{ni_2}^{a}} = \sum_{i_1i_2}\bra{\phi_{i_1}^{a}}\wh{O}\ket{\phi_{i_2}^a} \sum_n^\text{val} f_n P_{ni_1}^{a*}P_{ni_2}^a,
\end{equation}
and likewise for the smooth waves.
\par Introducing the Hermitian one-center density matrix 
\begin{equation}\label{eq: density matrix}
D_{i_1i_2}^a =\sum_n f_n P_{ni_1}^{a*} P_{ni_2}^{a} = \sum_n f_n
\braket{\s{\psi}_n}{\s{p}_{i_1}^a}
\braket{\s{p}_{i_2}^a}{\s{\psi}_n}.
\end{equation}
We conclude that for any local operator $\wh{O}$, the expectation value is
\begin{equation}\label{eq: local exp values}
  \langle \wh{O} \rangle = \sum_n^\text{val} f_n \bra{\s{\psi}_n}\wh{O}\ket{\s{\psi}_n} + \sum_a \sum_{i_1i_2} \left( \bra{\phi_{i_1}^{a}}\wh{O}\ket{\phi_{i_2}^a} - \bra{\s{\phi}_{i_1}^{a}}\wh{O}\ket{\s{\phi}_{i_2}^a} \right)D_{i_1i_2}^a + \sum_a \sum_\alpha^\text{core} \bra{\phi_\alpha^{a,\text{core}}}\wh{O}\ket{\phi_\alpha^{a,\text{core}}}.
\end{equation}

\subsection{Densities}\label{sec: densities}
The electron density is obviously a very important quantity in
DFT, as all observables in principle are calculated as functionals
of the density. In reality the kinetic energy is calculated as a
functional of the orbitals, and some specific exchange-correlation
functionals also rely on KS-orbitals rather then the density for
their evaluation, but these are still \emph{implicit}
functionals of the density.
\par To obtain the electron density we need to determine the
expectation value of the real-space projection operator
$\ket{\f{r}}\bra{\f{r}}$
\begin{equation}
n(\f{r}) = \sum_n f_n \braket{\psi_n}{\f{r}}\braket{\f{r}}{\psi_n} = \sum_n
f_n|\psi_n(\f{r})|^2,
\end{equation}
where $f_n$ are the occupation numbers.
\par As the real-space projection operator is obviously a local
operator, we can use the results \eqref{eq: local exp values} of the
previous section, and immediately arrive at
\begin{equation}\label{eq: electron density}
n(\f{r}) = \sum_n^\text{val} f_n |\s{\psi}_n|^2 + \sum_a
\sum_{i_1i_2} \left(\phi_{i_1}^{a} \phi_{i_2}^{a} -
\s{\phi}_{i_1}^{a} \s{\phi}_{i_2}^{a} \right) D_{i_1i_2}^a + \sum_a
\sum_\alpha^\text{core} |\phi_\alpha^{a,\text{core}}|^2.
\end{equation}
\par To ensure that \eqref{eq: electron density} reproduce the
correct density even though some of the core states are not
strictly localized within the augmentation spheres, a smooth core
density, $\s{n}_c(\f{r})$, is usually constructed, which is
identical to the core density outside the augmentation sphere, and
a smooth continuation inside. Thus the density is typically
evaluated as
\begin{equation}\label{eq: PAW density}
n(\f{r}) = \s{n}(\f{r}) + \sum_a \left( n^a(\f{r}) -
\s{n}^a(\f{r}) \right),
\end{equation}
where
\begin{subequations}\label{eq: density contributions}
\begin{align}
\s{n}(\f{r})   &= \sum_n^\text{val} f_n |\s{\psi}_n(\f{r})|^2 + \s{n}_c(\f{r})\label{eq: smooth n}\\
n^a(\f{r})     &= \sum_{i_1i_2} D_{i_1i_2}^a \phi_{i_1}^a(\f{r})\phi_{i_2}^{a}(\f{r}) + n_c^a(\f{r})\label{eq: partial n}\\
\s{n}^a(\f{r}) &= \sum_{i_1i_2} D_{i_1i_2}^a \s{\phi}_{i_1}^a(\f{r})\s{\phi}_{i_2}^{a}(\f{r}) + \s{n}_c^a(\f{r})\label{eq: smooth partial n}
\end{align}
\end{subequations}

\subsection{Total Energies}\label{sec: total energies}
The total energy of the electronic system is given by:
\begin{equation}
E[n] = T_s[n]+U_H[n]+V_{ext}[n]+E_{xc}[n].
\end{equation}
In this section, the usual energy expression above, is sought
re-expressed in terms of the PAW quantities: the smooth waves and the
auxiliary partial waves.
\par For the local and semi-local functionals, we can utilize
\eqref{eq: local exp values}, while the nonlocal parts needs more
careful consideration.

\subsubsection{The Semi-local Contributions}
\par The kinetic energy functional $T_s = \sum_n f_n \bra{\psi_n}
-\frac{1}{2}\nabla^2\ket{\psi_n}$ is obviously a (semi-) local
functional, so we can apply \eqref{eq: local exp values} and
immediately arrive at:
\begin{equation}
  \begin{split}
    T_s[\{\psi_n\}] &= \sum_n f_n \bra{\psi_n} -\tfrac{1}{2}\nabla^2\ket{\psi_n}\\
    &= \sum_n^\text{val} f_n \bra{\s{\psi}_n} - \tfrac{1}{2} \nabla^2\ket{\s{\psi}_n} + \sum_a \left(T_c^a + \Delta T_{i_1i_2}^a D^a_{i_1i_2} \right),
  \end{split}
\end{equation}
where
\begin{equation}
  T_c^a = \sum_\alpha^\text{core} \bra{\phi_\alpha^{a,\text{core}}} - \tfrac{1}{2}\nabla^2 \ket{\phi_\alpha^{a,\text{core}}} \quad \text{and} \quad \Delta T_{i_1i_2}^a = \bra{\phi_{i_1}^{a}} - \tfrac{1}{2}\nabla^2\ket{\phi_{i_2}^a} - \bra{\s{\phi}_{i_1}^{a}} - \tfrac{1}{2} \nabla^2 \ket{\s{\phi}_{i_2}^a}.
\end{equation}
For LDA and GGA type exchange-correlation functionals, $E_{xc}$ is
likewise, per definition, a semi-local functional, such that it can be
expressed as
\begin{equation}
  E_{xc}[n] =   E_{xc}[\s{n}] + \sum_a \left( E_{xc}[n^a] - E_{xc}[\s{n}^a] \right).
\end{equation}
By virtue of \eqref{eq: partial n}-\eqref{eq: smooth partial n} we can
consider the atomic corrections as functionals of the density matrix
defined in \eqref{eq: density matrix}, i.e.
\begin{equation}
  E_{xc}[n] =   E_{xc}[\s{n}] + \sum_a \Delta E_{xc}^a[\{D^a_{i_1i_2}\}],
\end{equation}
where
\begin{equation}
  \Delta E_{xc}^a[\{D^a_{i_1i_2}\}] = E_{xc}[n^a] - E_{xc}[\s{n}^a].
\end{equation}

\subsubsection{The Nonlocal Contributions}
The Hartree term is both nonlinear and nonlocal, so more care needs to be taken when introducing the PAW transformation for this expression.
\par In the following we will assume that there is no `true' external field, such that $V_\text{ext}[n]$ is only due to the static nuclei, i.e. it is a sum of the classical interaction of the electron density with the static ionic potential, and the electrostatic energy of the nuclei.
\par We define the total classical electrostatic energy functional as
\begin{equation}\label{eq: coulomb energy}
  \begin{split}
    E_C[n] &= U_H[n] + V_\text{ext}[n] = \frac{1}{2} ((n)) + (n|\textstyle\sum_a Z^a) + \frac{1}{2} \sum_{a\neq a'} (Z^a | Z^{a'}),
  \end{split}
\end{equation}
where the notation (f|g) indicates the Coulomb integral
\begin{equation}
(f|g) = \iint d\br d\br' \frac{f^*(\br) g(\br') }{|\br-\br'|}
\end{equation}
and I have introduced the short hand notation $((f)) = (f|f)$. In \eqref{eq: coulomb energy}, $Z^a(\br)$ is the charge density of the nucleus at atomic site $a$, which in the classical point approximation is given by
\begin{equation}
  Z^a(\br) = -\Z^a\delta(\br-\bR^a)
\end{equation}
with $\Z^a$ being the atomic number of the nuclei. As the Hartree energy of a density with non-zero total charge is numerically inconvenient, we introduce the charge neutral total density
\begin{equation}
  \rho(\br) = n(\br) + \sum_a Z^a(\br) \quad (= n_\text{electrons} + n_\text{nuclei}).
\end{equation}
In terms of this, the coulombic energy of the system can be expressed by
\begin{equation}\label{eq: coulomb energy reduced}
  E_C[n] = U_H'[\rho] = \frac{1}{2}((n+{\textstyle\sum_a Z^a}))'
\end{equation}
where the prime indicates that one should remember the
self-interaction error of the nuclei introduced in the Hartree energy
of the total density. This correction is obviously ill defined, and
different schemes exist for making this correction. As it turns out,
this correction is handled very naturally in the PAW formalism.
\par For now, we will focus on the term $((\rho)) =
((n+\textstyle\sum_a Z^a))$. If one where to directly include the
expansion of $n(\br)$ according to \eqref{eq: PAW density}, one would
get:
\begin{align*}
  ((n+\textstyle\sum_a Z^a)) &= ((\s{n}+\textstyle\sum_a n^a - \s{n}^a + Z^a)) \\&= ((\s{n})) + \sum_{aa'}(n^a - \s{n}^a + Z^a|n^{a'} - \s{n}^{a'} + Z^{a'}) + 2\sum_a(\s{n}|n^a - \s{n}^a + Z^a),
\end{align*}
where in the last expression, the first term is the Hartree energy of
the smooth electron density, which is numerically problematic because
of the nonzero total charge. The second term contains a double
summation over all nuclei, which would scale badly with system size,
and the last term involves integrations of densities represented on
incompatible grids (remember that the one-center densities are
represented on radial grids to capture the oscillatory behavior near
the nuclei)\footnote{One could separate the terms in other ways, but
  it is impossible to separate the smooth and the localized terms
  completely.}. This is clearly not a feasible procedure. To correct
these problem we add and subtract some atom centered compensation
charges $\s{Z}^a$:
\begin{multline*}
  ((n+\textstyle\sum_a \s{Z}^a + \textstyle\sum_a \left[Z^a - \s{Z}^a\right])) = ((\s{n} + \textstyle\sum_a \s{Z}^a)) + \sum_{aa'}(n^a - \s{n}^a + Z^a - \s{Z}^a|n^{a'} - \s{n}^{a'} + Z^{a'}- \s{Z}^a) \\+ 2\sum_a(\s{n}+\textstyle\sum_{a'}\s{Z}^{a'}|n^a - \s{n}^a + Z^a - \s{Z}^a).
\end{multline*}
If we define $\s{Z}^a(\br)$ in such a way that $n^a(\br) -
\s{n}^a(\br) + Z^a(\br) - \s{Z}^a(\br)$ has no multipole moments, i.e.
\begin{equation}\label{eq: no multipole}
  \int d\br r^l Y_L(\wh{\br-\bR^a}) (n^a - \s{n}^a + Z^a - \s{Z}^a) = 0
\end{equation}
for all $a$, the potentials of these densities are zero outside their
respective augmentation spheres ($L=(l,m)$ is a collective angular-
and magnetic quantum number). Exploiting this feature, the Coulomb
integral reduce to
\begin{align*}
  ((n+\textstyle\sum_a Z^a)) 
  &= ((\s{n} + \textstyle\sum_a \s{Z}^a)) + \sum_{a}((n^a - \s{n}^a + Z^a - \s{Z}^a)) + 2\sum_a(\s{n}^a+\s{Z}^a|n^a - \s{n}^a + Z^a - \s{Z}^a)\\
  &= ((\s{n} + \textstyle\sum_a \s{Z}^a)) + \sum_{a}\left( ((n^a + Z^a)) - ((\s{n}^a + \s{Z}^a)) \right)
\end{align*}
where it has been used that inside the augmentation spheres $\s{n} =
\s{n}^a$. In this expression, we have circumvented all of the previous
problems. None of the terms correlates functions on different grids,
there is only a single summation over the atomic sites, and
furthermore the only thing that has to be evaluated in the full space
is the Hartree energy of $\s{n}(\br) + \sum_a \s{Z}^a(\br)$ which is
charge neutral (see eq. \eqref{eq: rhot charge neutral}).
\par Inserting the final expression in \eqref{eq: coulomb energy}, we see that
\begin{equation}
  \begin{split}
       E_C[n] &=  \frac{1}{2}((\s{n} + {\textstyle\sum_a} \s{Z}^a)) + \frac{1}{2}\sum_a \left(((n^a + Z^a))' - ((\s{n}^a + \s{Z}^a))\right)\\
              &=U_H[\s{\rho}] + \frac{1}{2}\sum_a \left( ((n^a)) + 2(n^a|Z^a) - ((\s{n}^a + \s{Z}^a))\right)
  \end{split}
\end{equation}
where we have introduced the smooth total density 
\begin{equation}
  \s{\rho}(\br) = \s{n} + \sum_a \s{Z}^a(\br).
\end{equation}
Note that the problem with the self interaction error of the nuclei
could easily be resolved once it was moved to the atom centered part,
as handling charged densities is not a problem on radial grids.
\par To obtain an explicit expression for the compensation charges, we
make a multipole expansion of $\s{Z}^a(\br)$
\begin{equation}\label{eq: compensation expansion}
  \s{Z}^a = \sum_L Q_L^a ~\s{g}_L^a(\br),
\end{equation}
where $\s{g}_L^a(\br)$ is any smooth function localized within
$|\br-\bR^a|<r_c^a$, satisfying
\begin{equation}
  \int d\br r^l Y_L(\wh{\br-\bR^a}) \s{g}_{L'}^a(\br) = \delta_{LL'}.
\end{equation}
%
\par Plugging the expansion \eqref{eq: compensation expansion} into
equations \eqref{eq: no multipole}, we see that the expansion
coefficients $Q_L^a$ from must be chosen according to
\begin{equation}\label{eq: compensation multipoles}
  Q_L^a = \int d\br r^l Y_L(\hat{\br}) \left(n^a(\br) - \s{n}^a(\br) + Z^a(\br)\right) =\Delta^a\delta_{l,0} + \sum_{i_1i_2}\Delta_{L i_1i_2}^aD_{i_1i_2}^a
\end{equation}
where
\begin{subequations}
  \begin{align}
    \Delta^a &= \int dr \left(n_c^a(r) - \s{n}_c^a(r)\right) - \Z^a / \sqrt{4\pi}\label{eq: Delta^a}\\
    \Delta_{L i_1i_2}^a &= \int d\br r^l Y_L(\h{\br})[\phi_{i_1}^a(\br)\phi_{i_2}^{a}(\br) - \s{\phi}_{i_1}^a(\br)\s{\phi}_{i_2}^{a}(\br)]\label{eq: Delta_Lij}
  \end{align}
\end{subequations}
and it has been used that the core densities are spherical $n_c^a(\br)
= n_c^a(r) Y_{00}(\hat{\br})$ (we consider only closed shell frozen
cores). This completely defines the compensation charges
$\s{Z}^a(\br)$.
\par Note that the special case $l=0$ of \eqref{eq: no multipole}, implies that
\begin{align}
\int d\f{r} &\left[ n^a - \s{n}^a +Z^a - \s{Z}^a \right] = 0\nonumber\\
& \Downarrow\nonumber\\
\int d\f{r} &\left[\s{n}(\br) + \sum_a \s{Z}^a(\br) \right] = \int d\f{r} \left[n(\br) + \sum_a Z^a(\br)\right]\nonumber\\
& \Updownarrow\nonumber\\
\int d\br &~\s{\rho}(\br) = \int d\br ~\rho(\br) = 0\label{eq: rhot charge neutral}
\end{align}
i.e. that the smooth total density has zero total charge, making the
evaluation of the Hartree energy numerically convenient.
\par In summary, we conclude that the classical coulomb interaction
energy which in the KS formalism was given by $E_C[n] = U_H'[\rho]$,
in the PAW formalism becomes a pure Hartree energy (no
self-interaction correction) of the smooth total density $\s{\rho}$
plus some one-center corrections
\begin{equation}
   E_C[n] = U_H[\s{\rho}] + \sum_a \Delta E_C^a[\{D^a_{i_1i_2}\}]
\end{equation}
where the corrections
\begin{align*}
  \Delta E_C^a[\{D^a_{i_1i_2}\}] &= \tfrac{1}{2} ((n^a)) + (n^a|Z^a) - \tfrac{1}{2} ((\s{n}^a)) - (\s{n}^a| \s{Z}^a) - \tfrac{1}{2} ((\s{Z}^a))\\
  &= \tfrac{1}{2} \left[((n_c^a)) - ((\s{n}_c^a))\right] - \Z^a\int d\br \frac{n_c^a(r)}{r} - \sum_L Q_L^a(\s{n}_c^a|\s{g}_L^a)\\
  &\quad+ \sum_{i_1i_2} D^{a*}_{i_1i_2}\left[ (\phi_{i_1}^a\phi_{i_2}^{a}|n_c^a) - (\s{\phi}_{i_1}^a\s{\phi}_{i_2}^{a}|\s{n}_c^a) - \Z^a\int d\br \frac{\phi_{i_1}^{a}(\br)\phi_{i_2}^{a}(\br)}{r} -\sum_L Q_L^a(\s{\phi}_{i_1}^a\s{\phi}_{i_2}^{a}|\s{g}_L^a)\right] \\
  &\quad+ \frac{1}{2}\sum_{i_1i_2i_3i_4} D^{a*}_{i_1i_2}\left[(\phi_{i_1}^a\phi_{i_2}^{a}|\phi_{i_3}^a\phi_{i_4}^{a}) - (\s{\phi}_{i_1}^a\s{\phi}_{i_2}^{a}|\s{\phi}_{i_3}^a\s{\phi}_{i_4}^{a}) \right]D^{a}_{i_3i_4} - \frac{1}{2}\sum_{LL'}Q_L^aQ_{L'}^a (\s{g}_L^a|\s{g}_{L'}^a)
\end{align*}
Using that the potential of a spherical harmonic (times some radial
function) is itself a spheical harmonic of the same angular momentum,
we see that $(\s{g}_L^a|\s{g}_{L'}^a) \propto \delta_{LL'}$ and
$(\s{n}_c^a|\s{g}_L^a) \propto \delta_{L0}$. Noting that $Q_L^a$ by
virtue of \eqref{eq: compensation multipoles} is a functional of the
density matrix, and inserting this, we get
\begin{equation}
  \Delta E_C^a = \Delta C^a + \sum_{i_1i_2} \Delta C^a_{i_1i_2} D^a_{i_1i_2} + \sum_{i_1i_2i_3i_4} D^{a*}_{i_1i_2} \Delta C^a_{i_1i_2i_3i_4} D^a_{i_3i_4}
\end{equation}
where
\begin{align}
  \Delta C^a =& \tfrac{1}{2} \left[((n_c^a)) - ((\s{n}_c^a)) - (\Delta^a)^2 ((\s{g}^a_{00}))\right] - \Delta^a (\s{n}_c^a|\s{g}_{00}^a) - \sqrt{4\pi}\Z^a \int dr \frac{n_c^a(r)}{r}\label{eq:coulom tensor 0}\\
  \Delta C^a_{i_1i_2} =& (\phi_{i_1}^a\phi_{i_2}^{a}|n_c^a) - (\s{\phi}_{i_1}^a\s{\phi}_{i_2}^{a}|\s{n}_c^a) - \Z^a\int d\br \frac{\phi_{i_1}^{a}(\br)\phi_{i_2}^{a}(\br)}{r} \nonumber \\
  &- \Delta^a (\s{\phi}_{i_1}^a\s{\phi}_{i_2}^{a}|\s{g}_{00}^a) - \Delta^a_{00,i_1i_2}\Big( \Delta^a (\s{n}_c^a|\s{g}_{00}^a) + ((\s{g}^a_{00})) \Big)\label{eq:coulom tensor 1}\\
  \Delta C^a_{i_1i_2i_3i_4} =& \tfrac{1}{2} \left[(\phi_{i_1}^a\phi_{i_2}^{a} | \phi_{i_3}^a\phi_{i_4}^{a}) - (\s{\phi}_{i_1}^a\s{\phi}_{i_2}^{a}|\s{\phi}_{i_3}^a\s{\phi}_{i_4}^{a}) \right] \nonumber\\
  & -\sum_L \left[ \tfrac{1}{2}\Delta^a_{Li_1i_2}(\s{\phi}_{i_1}^a\s{\phi}_{i_2}^{a}|\s{g}_L^a) + \tfrac{1}{2}\Delta^a_{Li_3i_4}(\s{\phi}_{i_3}^a\s{\phi}_{i_4}^{a}|\s{g}_L^a) + \Delta^a_{Li_1i_2}((\s{g}^a_L))\Delta^a_{Li_3i_4} \right]\label{eq:coulom tensor 2}
\end{align}
Note that all integrals can be limited to the inside of the
augmentation sphere. For example $(\phi_{i_1}^a\phi_{i_2}^{a}|n_c^a)$
has contributions outside the augmentation sphere, but these are
exactly canceled by the contributions outside the spheres of
$(\s{\phi}_{i_1}^a\s{\phi}_{i_2}^{a}|\s{n}_c^a)$, in which region the
two expressions are identical.
\par The $\Delta C^a_{i_1i_2i_3i_4}$ tensor has been written in a symmetric
form, such that it is invariant under the following symmetry
operations:
\begin{align}\label{eq: C symmetry}
  i_1 &\leftrightarrow i_2 &   i_3 &\leftrightarrow i_4 &   i_1i_2 &\leftrightarrow i_3i_4
\end{align}
%
%
%

\subsubsection{Summary}
Summing up all the energy contributions, we see that the Kohn-Sham total energy
\begin{equation*}
  E[n] = T_s[\{\psi_n\}] + U_H'[\rho] + E_{xc}[n]
\end{equation*}
can be separated into a part calculated on smooth functions, $\s{E}$,
and some atomic corrections, $\Delta E^a$, involving quantities
localized around the nuclei only.
\begin{equation}\label{eq: PAW energy functional}
E = \s{E} + \sum_a \Delta E^a
\end{equation}
where the smooth part
\begin{equation}
  \s{E} = T_s[\{\s{\psi}_n\}] + U_H[\s{\rho}] + E_{xc}[\s{n}]
\end{equation}
is the usual energy functional, but evaluated on the smooth functions
$\s{n}$ and $\s{\rho}$ instead of $n$ and $\rho$, and with the soft
compensation charges $\s{Z}^a$ instead of the nuclei charges
$Z^a(\br)$. The corrections are given by
\begin{equation}
  \Delta E^a = \Delta T_c^a + \Delta C^a + \sum_{i_1i_2} \left(\Delta T^a_{i_1i_2} + \Delta C^a_{i_1i_2}\right) + \sum_{i_1i_2i_3i_4} D^{a*}_{i_1i_2} \Delta C^a_{i_1i_2i_3i_4} D^a_{i_3i_4} + \Delta E_{xc}^a[\{D^a_{i_1i_2}\}]
\end{equation}
where $T^a_c$, $\Delta T_{i_1i_2}^a$, $\Delta C_{i_1i_2}^a$, and
$\Delta C_{i_1i_2i_3i_4}^a$ are system independent tensors that can be
pre-calculated and stored for each specie in the periodic table of
elements.
\par Both the Hamiltonian and the forces can be derived from the total
energy functional \eqref{eq: PAW energy functional} as will be shown
in the following two sections.

\subsection{The Transformed Kohn-Sham Equation}
The variational quantity in the PAW formalism is the smooth wave
function $\s{\psi}_n$. From this, all other quantities, such as the
density matrix, the soft compensation charges, the transformation
operator, etc. are determined by various projections of $\s{\psi}_n$
onto the projector functions, and expansions in our chosen basis
functions, the partial and smooth partial waves. To obtain the smooth
wave functions, we need to solve the eigenvalue equation
\begin{equation}\label{eq: transformed KS}
  \wh{\ws{H}} \s{\psi}_n(\br) = \epsilon_n \h{S}\s{\psi}_n(\br),
\end{equation}
where the overlap operator $\h{S} = \T^\dagger \T$ and $\wh{\ws{H}} =
\T^\dagger \Ham \T$ is the transformed Hamiltonian.

\subsubsection{Orthogonality}
In the original form, the eigen states of the KS equation where
orthogonal, i.e. $\braket{\psi_n}{\psi_m} = \delta_{nm}$ while in the
transformed version
\begin{equation}
  \label{eq: orthogonality}
  \bra{\s{\psi}_n}\T^\dagger \T\ket{\s{\psi}_m} = \delta_{nm}
\end{equation}
i.e. the smooth wave function are only orthogonal with respect to the
weight $\h{S}$.
\par The explicit form of the overlap operator is
\begin{equation}\label{eq: overlap}
  \begin{split}
  \h{S} &= \T^\dagger \T \\
        &= \left(1+\textstyle\sum_a \T^a\right)^\dagger \left(1+\textstyle\sum_a \T^a\right)\\
        &= {1} + \sum_a \left( \T^{a\dagger} + \T^a + \T^{a\dagger}\T^a\right)\\
        &= {1} + \sum_a \Big[\sum_{i_1} \ket{\s{p}_{i_1}^a}(\bra{\phi_{i_1}^a} - \bra{\s{\phi}_{i_1}^a}) \sum_{i_2} \ket{\s{\phi}_{i_2}^a}\bra{\s{p}_{i_2}^a} + \sum_{i_2} \ket{\s{\phi}_{i_2}^a}\bra{\s{p}_{i_2}^a} \sum_{i_1}(\ket{\phi_{i_1}^a} - \ket{\s{\phi}_{i_1}^a})\bra{\s{p}_{i_1}^a} \\
        &\hspace*{20mm} + \sum_{i_1} \ket{\s{p}_{i_1}^a}(\bra{\phi_{i_1}^a} - \bra{\s{\phi}_{i_1}^a})\sum_{i_2}(\ket{\phi_{i_2}^a} - \ket{\s{\phi}_{i_2}^a})\bra{\s{p}_{i_2}^a}\Big]\\
        &= {1} + \sum_a \sum_{i_1i_2} \ket{\s{p}_{i_1}^a}(\braket{\phi_{i_1}^a}{\phi_{i_2}^a} - \braket{\s{\phi}_{i_1}^a}{\s{\phi}_{i_2}^a})\bra{\s{p}_{i_2}^a}\\
        &= {1} + \sum_a \sum_{i_1i_2} \ket{\s{p}_{i_1}^a}\sqrt{4\pi}\Delta^a_{00,i_1i_2}\bra{\s{p}_{i_2}^a}
  \end{split}
\end{equation}
The orthogonality condition \eqref{eq: orthogonality} must be kept in
mind when applying numerical schemes for solving \eqref{eq:
  transformed KS}. For example plane waves are no longer orthogonal,
in the sense that $\bra{\f{G}} \h{S} \ket{\f{G}'} \neq \delta_{\f{G},
  \f{G}'}$.

\subsubsection{The Hamiltonian}
To determine the transformed Hamiltonian, one could apply the
transformation $\wh{\ws{H}} = \T^\dagger \Ham \T$ directly, which
would be straight forward for the local parts of $\Ham$, but to take
advantage of the trick used to determine the total energy of the
nonlocal term ($E_C[n]$), we make use of the relation
\begin{equation}
  \frac{\delta E}{\delta \s{\psi}^*_n(\br)} = f_n \wh{\ws{H}} \s{\psi}_n(\br).
\end{equation}
Using this, we get
\begin{align*}
\frac{\delta E}{\delta \s{\psi}^*_n(\br)} &= \frac{\delta}{\delta \s{\psi}^*_n(\br)} \left[ T_s[\{\s{\psi}_n\}] + E_{xc}[\s{n}] + U_H[\s{\rho}] + \Delta E^a[\{D^a_{i_1i_2}\}]\right]\\
&= \frac{\delta T_s[\{\s{\psi}_n\}]}{\delta \s{\psi}^*_n(\br)} \\
&\quad + \int d\br' \left[ \frac{\delta E_{xc}[\s{n}]}{\delta \s{n}(\br')} + \frac{\delta U_H[\s{\rho}]}{\delta \s{n}(\br')} \right] \frac{\delta \s{n}(\br')}{\delta \s{\psi}^*_n(\br)}\\
&\quad + \sum_a \sum_{i_1i_2} \left[\int d\br' \frac{\delta U_H[\s{n} + \textstyle\sum_a \s{Z}^a]}{\delta \s{Z}^a(\br')}\frac{\delta \s{Z}^a(\br')}{\delta D^a_{i_1i_2}} + \frac{\delta \Delta E^a}{\delta D^a_{i_1i_2}} \right]\frac{\delta D^a_{i_1i_2}}{\delta \s{\psi}^*_n(\br)}\\
&= f_n (-\tfrac{1}{2}\nabla^2)\psi_n(\br)\\
&\quad + \int d\br' \left[ v_{xc}[\s{n}](\br') + u_H[\s{\rho}](\br') \right] f_n \delta(\br-\br')\s{\psi}_n(\br')\\
&\quad + \sum_a \sum_{i_1i_2} \left[\int d\br' u_H[\s{n} + \textstyle\sum_a\s{Z}^a](\br')\sum_L \Delta^a_{Li_1i_2}\s{g}^a_L(\br') + \frac{\delta \Delta E^a}{\delta D^a_{i_1i_2}} \right] f_n\s{p}^a_{i_1}(\br)P^a_{ni_2}
\end{align*}
where $v_{xc}[n](\br) = \frac{\delta E_{xc}[n]}{\delta n(\br)}$ is the
usual local (LDA) or semilocal (GGA) exchange correlation potential,
and $u_H[n](\br) = \frac{\delta U_H[n]}{\delta n(\br)} = \int d\br'
\frac{n(\br')}{|\br-\br'|}$ is the usual Hartree potential.
\par From these results, we can write down the transformed Hamiltonian as
\begin{equation}
  \wh{\ws{H}} = -\tfrac{1}{2}\nabla^2 + u_H[\s{\rho}] + v_{xc}[\s{n}] + \sum_a \sum_{i_1i_2} \ket{\s{p}^a_{i_1}} \Delta H^a_{i_1i_2} \bra{\s{p}^a_{i_2}},
\end{equation}
where the nonlocal part of the Hamiltonian is given in terms of the tensor
\begin{equation}
  \begin{split}
    \Delta H^a_{i_1i_2} &= \sum_L \Delta^a_{Li_1i_2} \int d\br u_H[\s{\rho}](\br)\s{g}^a_L(\br) + \frac{\delta \Delta E^a}{\delta D_{i_1i_2}^a}\\
    &= \sum_L \Delta^a_{Li_1i_2} \int d\br u_H[\s{\rho}](\br)\s{g}^a_L(\br) + \Delta T^a_{i_1i_2} + \Delta C^a_{i_1i_2} + 2\sum_{i_3i_4} \Delta C^a_{i_1i_2i_3i_4}D^a_{i_3i_4} + \frac{\delta \Delta E_{xc}}{\delta D^a_{i_1i_2}} .
  \end{split}
\end{equation}
Note that to justify taking the derivative with respect to $D$ only,
and not its complex conjugate, the symmetry properties \eqref{eq: C
  symmetry} has been used to get $D^{a*}_{i_1i_2}
\Delta C^a_{i_1i_2i_3i_4}D^a_{i_3i_4} = D^{a}_{i_1i_2}
\Delta C^a_{i_1i_2i_3i_4}D^a_{i_3i_4}$.

\subsection{Forces in PAW}\label{sec: forces}
In the ground state, the forces on each nuclei can be calculated directly from
\begin{equation}
  \label{eq: force}
  \begin{split}
    \f{F}^a &= -\frac{dE}{d\bR^a} \\
&= -\frac{\partial E}{\partial \bR^a} - \sum_n\left\{ \frac{\partial E}{\partial \ket{\s{\psi}_n}} \frac{d \ket{\s{\psi}_n}}{d \bR^a} + h.c.\right\}\\
&= -\frac{\partial E}{\partial \bR^a} - \sum_n f_n \epsilon_n \left\{ \bra{\s{\psi}_n}\h{S} \frac{d \ket{\s{\psi}_n}}{d \bR^a} + h.c.\right\}\\
&= -\frac{\partial E}{\partial \bR^a} + \sum_n f_n \epsilon_n \bra{\s{\psi}_n}\frac{d\h{S}}{d\bR^a}\ket{\s{\psi}_n}
  \end{split}
\end{equation}
where $h.c.$ denotes the hermitian conjugate. To get to the second
line, the chain rule has been applied. The third line follows from the
relation
\begin{equation}
  \frac{\partial E}{\partial \bra{\s{\psi}_n}} = f_n \wh{\ws{H}} \ket{\s{\psi}_n} = f_n \epsilon_n \h{S} \ket{\s{\psi}_n}.
\end{equation}
The last line of \eqref{eq: force} is obtained from the following
manipulation of the orthogonality condition \eqref{eq: orthogonality}
\begin{equation}
  \begin{split}
    &\quad \delta_{nm} = \bra{\s{\psi}_n}\h{S}\ket{\s{\psi}_m} \\
    \Rightarrow &\quad 0 = \frac{d}{d\bR^a} \bra{\s{\psi}_n}\h{S}\ket{\s{\psi}_m} = \frac{d\bra{\s{\psi}_n}}{d\bR^a}\h{S}\ket{\s{\psi}_m}  + \bra{\s{\psi}_n}\frac{d\h{S}}{d\bR^a}\ket{\s{\psi}_m} + \bra{\s{\psi}_n}\h{S}\frac{d\ket{\s{\psi}_m}}{d\bR^a}\\
    \Leftrightarrow & \quad \frac{d\bra{\s{\psi}_n}}{d\bR^a}\h{S}\ket{\s{\psi}_m} + h.c. = - \bra{\s{\psi}_n}\frac{d\h{S}}{d\bR^a}\ket{\s{\psi}_m}
  \end{split}
\end{equation}
\par From the expression for the overlap operator \eqref{eq: overlap},
it follows that
\begin{equation}
  \frac{d\h{S}}{d\bR^a} = \sum_{i_1i_2} \Delta S^a_{i_1i_2}\left(\frac{d \ket{\s{p}^a_{i_1}}}{d\bR^a}\bra{\s{p}^a_{i_2}} + h.c. \right)
\end{equation}
which, when inserted in \eqref{eq: force}, gives the force expression
\begin{equation}
  \f{F}^a = -\frac{\partial E}{\partial \bR^a} + \sum_n f_n \epsilon_n \sum_{i_1i_2}\Delta S^a_{i_1i_2}\left( P^{a*}_{ni_1}\braket{\frac{d \s{p}^a_{i_2}}{d\bR^a}}{\s{\psi}_n} + \braket{\s{\psi}_n}{\frac{d \s{p}^a_{i_1}}{d\bR^a}} P^a_{ni_2}\right)
\end{equation}
In the case of standard xc approximations, the dependence of the total
energy on the nuclei coordinates is
\begin{equation}
  \begin{split}
    \frac{\partial E}{\partial \bR^a} &= \int d\br' \frac{\delta E}{\delta \s{n}(\br')}\frac{\partial \s{n}(\br')}{\partial \bR^a} + \sum_{i_1i_2} \frac{\partial E}{\partial D^a_{i_1i_2}}\frac{\partial D^a_{i_1i_2}}{\partial \bR^a} + \int d\br' \sum_L \frac{\delta E}{\delta \s{g}^a_L(\br')}\frac{\partial \s{g}^a_L(\br')}{\partial \bR^a}\\
    &=\int d\br' \s{v}_\text{eff}(\br')\frac{\partial \s{n}^a_c(\br')}{\partial \bR^a} + \sum_{i_1i_2} \Delta H^a_{i_1i_2}\frac{\partial D^a_{i_1i_2}}{\partial \bR^a} + \int d\br' \sum_L u_H(\br')Q_L\frac{\partial \s{g}^a_L(\br')}{\partial \bR^a}
  \end{split}
\end{equation}
giving the force expression
\begin{equation}\label{eq: force no exx}
  \begin{split}
    \f{F}^a = - &\int d\br' \left\{ \s{v}_\text{eff}(\br')\frac{\partial \s{n}^a_c(\br')}{\partial \bR^a} + u_H(\br')\sum_L Q_L\frac{\partial \s{g}^a_L(\br')}{\partial \bR^a} \right\}\\
    &- \sum_n f_n \sum_{i_1i_2}\left\{\Delta H^a_{i_1i_2} - \epsilon_n\Delta S^a_{i_1i_2}\right\}\left( P^{a*}_{ni_1}\braket{\frac{d \s{p}^a_{i_2}}{d\bR^a}}{\s{\psi}_n} + \braket{\s{\psi}_n}{\frac{d \s{p}^a_{i_1}}{d\bR^a}} P^a_{ni_2}\right)
  \end{split}
\end{equation}

\subsection{Summary}\label{sec: summary exx hamiltonian}
The PAW KS equation to be solved is
\begin{equation}\label{eq: PAW HF-KS EVP}
  \wh{\ws{H}} \ket{\s{\psi}_n}= \epsilon_n \h{S} \ket{\s{\psi}_n}
\end{equation}
with $\h{S}$, and $\wh{\ws{H}}$ given by
\begin{subequations}
\begin{align}
\wh{S} &= \h{1} + \sum_a \sum_{i_1i_2} \ket{\s{p}_{i_1}^a}\sqrt{4\pi}\Delta^a_{00,i_1i_2}\bra{\s{p}_{i_2}^a}\\
\wh{\ws{H}} &= -\tfrac{1}{2}\nabla^2 + u_H[\s{\rho}](\br) + v_{xc}[\s{n}](\br) + \sum_a \sum_{i_1i_2} \ket{\s{p}^a_{i_1}} \Delta H^a_{i_1i_2} \bra{\s{p}^a_{i_2}}\label{eq: PAW Hamiltonian summary}
\end{align}
\end{subequations}
where
\begin{equation}
    \Delta H^a_{i_1i_2} = \sum_L \Delta^a_{Li_1i_2} \int d\br u_H[\s{\rho}](\br)\s{g}^a_L(\br) + \Delta T^a_{i_1i_2} + \Delta C^a_{i_1i_2} + 2\sum_{i_3i_4} \Delta C^a_{i_1i_2i_3i_4}D^a_{i_3i_4} + \frac{\delta \Delta E_{xc}}{\delta D^a_{i_1i_2}}
\end{equation}
\par The total energy can then be evaluated by
\begin{equation}
  E = T_s[\{\s{\psi}_n\}] + U_H[\s{\rho}] + E_{xc}[\s{n}] + \sum_a \Delta E^a  
\end{equation}
with $\Delta E^a$ given by
\begin{equation}
 \Delta E^a = T_c^a + \sum_{i_1i_2} \left( \Delta T^a_{i_1i_2} + \Delta C^a_{i_1i_2}\right) D_{i_1i_2}^a + \sum_{i_1i_2i_3i_4}D_{i_1i_2}^{a*} \Delta C_{i_1i_2i_3i_4}^a D_{i_3i_4}^a + \Delta E_{xc}^a(\{D_{i_1i_2}^a\})
\end{equation}
\par Having solved the eigenvalue problem \eqref{eq: PAW HF-KS EVP},
the eigenvalues are known. This can be used to determine, for example,
the kinetic energy of the pseudo wave functions, $T_s[\s{n}]$, without
doing the explicit (and computationally costly) computation. This can
be seen by operating with $\sum_n f_n \bra{\s{\psi}_n}$ on eq.
\eqref{eq: PAW Hamiltonian summary} to get:
\begin{equation}
  \label{eq:kinetic-energy}
  T_s[\{\s{\psi}_n\}] = \sum_n f_n \epsilon_n - \int d\br [\s{n}(\br) - \s{n}_c(\br)] \left[u_H[\s{\rho}](\br) + v_{xc}[\s{n}](\br)\right] - \sum_a \sum_{i_1i_2} \Delta H^a_{i_1i_2} D^a_{i_1i_2}
\end{equation}

\section{Implementing PAW}\label{sec: implementing}
For an implementation of PAW, one must specify a large number of data
for each chemical element. This constitutes a data set which uniquely
determines how the on-site PAW transformation works, at the site of
the specific atom. For the generation of such data sets, one needs an
atomic DFT program, with which basis sets can be generated. How to
perform DFT calculations efficiently on an isolated atom will be
discussed in the first section of this chapter, and the actual choice
of data set parameters will be discussed in the second. The atomic DFT
program plays the additional role of a small test program, against
which implementations in the full PAW program can be tested.

\subsection{Atoms}
If we consider the Kohn-Sham equation for an isolated atom, (described
by a non spin-dependent Hamiltonian), it is well known that the
eigenstates can be represented by the product
\begin{equation}
  \phi_{i\sigma_i}(\br\sigma) = R_j(r) Y_L(\h{\br})\chi_{\sigma_i}(\sigma)
\end{equation}
where $R_j$ are real radial function, and $Y_L$ are the (complex
valued) spherical harmonics. $i=(n,l,m)$, $j=(n,l)$, and $L = (l,m)$.
\par Assuming identical filling of all atomic orbitals, i.e. $f_{i
  \sigma} = f_j$, the density becomes
\begin{equation}
  n(\br) = \sum_i \sum_{\sigma_i} f_j|\phi_{i\sigma_i}(\br\sigma)|^2 = \sum_j 2 \frac{2l+1}{4\pi} f_j|R_j(r)|^2
\end{equation}
where the first factor of 2 comes from the sum over spin, and the
second factor from the sum over the magnetic quantum number using that
\begin{equation}
  \sum_m |Y_{lm}|^2 = \frac{2l+1}{4\pi}
\end{equation}
\par The identical filling of degenerate states is exact for closed
shell systems, and corresponds to a spherical averaging of the density
for open shell systems.
\par Determining potentials in a spherical coordinate system is
usually done by exploiting the expansion of the Coulomb kernel
\begin{equation}
  \frac{1}{|\br-\br'|} = \sum_L \frac{4\pi}{2l+1} \frac{r_<^l}{r_>^{l+1}} Y_L^*(\h{\br})Y_L(\h{\br}')
\end{equation}
with $r_< = \min(r, r')$ and $r_> = \max(r, r')$. Using this it is
seen that for any density with a known angular dependence, e.g. the
density $R(r) Y_L(\h{\br})$, the potential can be determined by
\begin{equation}\label{eq: radial potential}
  \begin{split}
    v[R(r) Y_L(\h{\br})](\br) &= \int d\br' \frac{R(r') Y_L(\h{\br}')}{|\br - \br'|}\\
    &= \frac{4\pi}{2l+1} Y_L(\h{\br}) \int_0^\infty r'^2dr' R(r') \frac{r_<^l}{r_>^{l+1}}\\
    &= \frac{4\pi}{2l+1} Y_L(\h{\br}) \left[\int_0^r dr' R(r')r'\Big(\frac{r'}{r}\Big)^{l+1} + \int_r^\infty dr' R(r')r'\Big(\frac{r}{r'}\Big)^l \right]
  \end{split}
\end{equation}
if the angular dependence is not a spherical harmonic, one can always
do a multipole expansion, and use the above expression on the
individual terms.
\par In the case of a radial density $n(\br) = n(r)$, the Hartree
potential becomes
\begin{equation}
  u_H(r) = \frac{4\pi}{r}\int_0^r dr' n(r')r'^2 + 4\pi\int_r^\infty dr' n(r')r'
\end{equation}
A purely radial dependent density also implies that the xc-potential
is a radial function. Using this, the entire KS equation can be
reduced to a 1D problem in $r$, while the angular part is treated
analytically.

\subsubsection{The Radial Kohn-Sham Equation}
For a spherical KS potential, and using that $Y_L$ are eigenstates of
the Laplacian, the KS equation can be reduced to the simpler
one-dimensional second order eigenvalue problem
\begin{equation}\label{eq: radial KS equation}
  \left[ -\frac{1}{2}\frac{d^2}{dr^2} - \frac{1}{r}\frac{d}{dr} + \frac{l(l+1)}{2r^2} + v_s(r)\right]R_j(r) = \epsilon_j R_j(r)
\end{equation}
If we introduce the radial wave function $u_j(r)$ defined by
\begin{equation}
  r R_j(r) = u_j(r)
\end{equation}
the KS equation can be written as
\begin{equation}
  u_j''(r)  + \left(2\epsilon_j - 2v_s(r)- \frac{l(l+1)}{r^2}\right) u_j(r) = 0
\end{equation}
which is easily integrated using standard techniques. See e.g. \cite[chapter 6]{Fiolhais2003}.

\subsection{The Atomic Data Set of PAW}\label{sec: partial wave basis}
The very large degree of freedom when choosing the functions defining
the PAW transformation means that the choice varies a great deal
between different implementations. In any actual implementation
expansions are obviously finite, and many numerical considerations
must be made when choosing these basis sets, to ensure fast and
reliable convergence. This section provides an overview of the
information needed for uniquely defining the PAW transformation, and
the level of freedom when choosing the parameters.

\subsubsection*{The Partial Waves}
The basis functions, $\phi_i^a$, for the expansion of
$\ket{\psi_n}$ should be chosen to ensure a fast convergence to
the KS wave function. For this reason we choose the partial waves
as the eigenstates of the Schrödinger equation for the isolated
spin-saturated atoms. Thus the index $i$ is a combination of
main-, angular-, and magnetic quantum number, $(n,l,m)$. And the
explicit form is
\begin{equation*}
\phi_i^a(\f{r})=\phi_{nl}^a(r)Y_{lm}(\hat{\f{r}})
\end{equation*}
where $\phi_{nl}^a(r)$ are the solutions of the radial KS Schrödinger
equation \eqref{eq: radial KS equation}, and $Y_{lm}$ are the
spherical harmonics. For convenience we choose $\phi_i^a(\f{r})$ to be
real, i.e. we use the real spherical harmonics instead of the complex
valued. This choice of partial waves implies that the smooth partial
waves and the smooth projector functions can also be chosen real, and
as products of some radial functions and the same real spherical
harmonic.
\par Note that including unbound states of the radial KS equation in
the partial waves is not a problem, as the diverging tail is exactly
canceled by the smooth partial waves. In practice we only integrate
the KS equation outward from the origin to the cutoff radius for
unbound states, thus making the energies free parameters. In principle
the same could be done for the bound states, but in \gpaw, the
energies of bound states are fixed by making the inward integration
for these states and doing the usual matching (see e.g. \cite[chapter
6]{Fiolhais2003}), i.e. the energies are chosen as the eigen energies
of the system.

\subsubsection*{The Smooth Partial Waves}
\par The smooth partial waves $\s{\psi}_i^a(\br)$ are per construction
identical to the partial waves outside the augmentation sphere. Inside
the spheres, we can choose them as any smooth continuation. Presently
\gpaw{} uses simple 6'th order polynomials of even powers only (as odd
powers in $r$ results in a kink in the functions at the origin, i.e.
that the first derivatives are not defined at this point), where the
coefficients are used to match the partial waves smoothly at $r=r_c$.
Other codes uses Bessel functions\cite{Kresse1999} or Gaussians.

\subsubsection*{The Smooth Projector Functions}
\par The smooth projector functions are a bit more tricky.  Making
them orthonormal to $\s{\phi}_i^a(\f{r})$ is a simple task of applying
an orthonormalization procedure. This is the only formal requirement,
but in any actual implementation all expansions are necessarily
finite, and we therefore want them to converge as fast as possible, so
only a few terms needs to be evaluated.
\par A popular choice is to determine the smooth projector functions
according to
\begin{equation}\label{eq: construct projector}
  \ket{\s{p}_i^a} = \left( -\tfrac{1}{2} \nabla^2 + \s{v}_s - \epsilon_i\right) \ket{\s{\phi}^a_i}
\end{equation}
or equivalently
\begin{equation}
  \s{p}_{j}^a(r) = \left[-\frac{1}{2}\frac{d^2}{dr^2} - \frac{1}{r}\frac{d}{dr} + \frac{l(l+1)}{2r^2} + \s{v}_s(r) - \epsilon_j \right] \s{\phi}^a_j(r)
\end{equation}
where $\s{v}_s(r)$ is the smooth KS potential $u_H[\s{\rho}](r) +
v_{xc}[\s{n}](r)$. And then enforce the complementary orthogonality
condition $\braket{\s{p}_{j}^a}{\s{\phi}^a_{j'}} = \delta_{j,j'}$
inside the augmentation sphere, e.g. by a standard Gram-Schmidt
procedure. Using this procedure ensures that the reference atom is
described correctly despite the finite number of projectors.

\subsubsection*{The Smooth Compensation Charge Expansion Functions}\label{sec: choosing comp charge}
The smooth compensation charges $\s{g}_L^a(\br)$, are products of
spherical harmonics, and radial functions $\s{g}_l^a(r)$ satisfying
that
\begin{equation}
\int d\f{r} r^l Y_L(\hat{\f{r}})\s{g}_{L'}^a(\f{r}) = \delta_{LL'}
\end{equation}
In \gpaw{} the radial functions are chosen as generalized Gaussian
according to (here shown for $\bR^a=0$):
\begin{equation}\label{eq: generalized gaussians}
\s{g}_L^a(\f{r}) = \s{g}_l^a(r) Y_L(\hat{\f{r}})~~,\quad
\s{g}_l^a(r) = \frac{1}{\sqrt{4\pi}}\frac{l!}{(2l+1)!}(4\alpha^a)^{l+3/2}r^le^{-\alpha^ar^2}
\end{equation}
where the atom-dependent decay factor $\alpha$ is chosen such that
the charges are localized within the augmentation sphere.

\subsubsection*{The Core- and Smooth Core Densities }
The core density follows directly from the all electron partial waves by
\begin{equation}\label{eq: core density}
n_c(r) = \sum_i^\text{core} |\phi_i(\br)|^2 = \sum_j^\text{core} 2(2l+1) |\phi_j(r)|^2 / 4\pi
\end{equation}
\par The smooth core densities $\s{n}_c^a(\br)$ are like the smooth
partial waves expanded in a few (two or three) Bessel functions,
Gaussians, polynomials or otherwise, fitted such that it matches the
true core density smoothly at the cut-off radius.

\subsubsection*{The Localized Potential}
An additional freedom in PAW is that for any operator $\wh{L}$,
localized within the augmentation spheres, we can exploit the identity
\eqref{eq: phi p completeness}
\begin{equation}
\sum_i \ket{\s{\phi}_i^a}\bra{\s{p}_i^a} = 1 
\end{equation}
valid within the spheres, to get
\begin{equation*}
  \wh{L} = \sum_a \sum_{i_1i_2} \ket{\s{p}^a_{i_1}}\bra{\s{\phi}_{i_1}^a}\wh{L} \ket{\s{\phi}_{i_2}^a}\bra{\s{p}^a_{i_2}}
\end{equation*}
so for any potential $\bar{v}(\br) = \sum_a \bar{v}^a(\br- \bR^a)$
localized within the augmentation spheres (i.e. $\bar{v}^a(\br) = 0$
for $r>r_c^a$) we get the identity
\begin{equation*}
  0 = \int d\br \s{n}(\br) \sum_a \bar{v}^a(\br) - \sum_a \int d\br \s{n}^a \bar{v}^a(\br)
\end{equation*}
This expression can be used as an `intelligent zero'. Using this, we
can make the replacement of the smooth potential
\begin{equation}
  \s{v}_s(\br) = u_H[\s{\rho}](\br) + v_{xc}[\s{n}](\br) \to \s{v}_s(\br) = u_H[\s{\rho}](\br) + v_{xc}[\s{n}](\br) + \bar{v}(\br)  
\end{equation}
if we at the same time add 
\begin{equation}
  B^a + \sum_{i_1i_2} \Delta B^a_{i_1i_2} D^a_{i_1i_2}
\end{equation}
to the energy corrections $\Delta E^a$, where
\begin{equation}
  B^a = -\int d\br \s{n}_c^a\bar{v}^a(\br) \quad\text{and}\quad \Delta B^a_{i_1i_2} = -\int d\br \s{\phi}^a_{i_1}\s{\phi}^a_{i_2}\bar{v}^a(\br)
\end{equation}
This also implies that $B^a_{i_1i_2}$ should be added to $\Delta
H^a_{i_1i_2}$.
\par The advantage of doing this is that the Hartree potential and the
xc-potential might not be optimally smooth close to the nuclei, but if
we define the localized potential properly, the sum of the three
potentials might still be smooth. Thus one can initially evaluate
$u_H[\s{\rho}](\br)$ and $v_{xc}[\s{n}](\br)$ on an extra fine grid,
add $\bar{v}(\br)$ and then restrict the total potential to the coarse
grid again before solving the KS equation.
\par The typical way of constructing the localized potentials
$\bar{v}^a$ is by expanding it in some basis, and then choosing the
coefficients such that the potential $u_H[\s{\rho}](\br) +
v_{xc}[\s{n}](\br) + \bar{v}(\br)$ is optimally smooth at the core for
the reference system.
\par Inclusion of $\bar{v}^a(\br)$ changes the forces on each atom only through the redefinitions of $\s{v}_s(\br)$ and $\Delta H^a_{i_1i_2}$.

\subsubsection*{Summary}
When constructing a data set for a specific atom, one must specify the
following quantities, all defined within the augmentation spheres
only:
\begin{enumerate}
\item $\phi_i^a$ from radial KS equation
\item $\s{\phi}_i^a$ by appropriate smooth continuation of $\phi_i^a$
\item $\s{p}_i^a$ from equation \eqref{eq: construct projector}
\item $\s{g}_L^a$ localized within $r<r_c$, and satisfying $\int d\f{r} r^{l'} \s{g}_L^a(\f{r})Y_{L'}(\wh{\br - \bR^a}) = \delta_{LL'}$
\item $n_c^a$ follows from $\phi_i^a$ by \eqref{eq: core density}
\item $\s{n}_c^a$ by appropriate smooth continuation of $n_c^a$
\item $\bar{v}^a$ localized within $r<r_c^a$, otherwise freely chosen to make $\s{v}_s$ optimally smooth for the reference system
\end{enumerate}
The adjustable parameters besides the shape of $\s{\phi}^a$,
$\s{g}_L^a$, $\bar{v}^a$, and $\s{n}_c^a$ are
\begin{enumerate}
\item Cut-off radii $r_c^a$ (which can also depend on $i$)
\item Frozen core states
\item Number of basis set functions (range of index $i$ on $\phi_i^a$,
  $\s{\phi}_i^a$, and $\s{p}_i^a$)
\item Energies of unbound partial waves
\end{enumerate}
Choosing these parameters in such a way that the basis is sufficient
for the description of all possible environments for the specific
chemical element, while still ensuring a smooth pseudo wave function
is a delicate procedure, although the optimal parameter choice is more
stable than for e.g. norm conserving or ultra soft pseudopotentials.
\par Once the quantities above have been constructed, all other
ingredients of the PAW transformation follows, such as $\Delta^a$,
$\Delta_{Lii'}^a$, $T_c^a$, $\Delta T^a_{i_1i_2}$, $\Delta C^a$,
$\Delta C^a_{i_1i_2}$, $\Delta C^a_{i_1i_2i_3i_4}$, $\Delta
\bar{v}^a$, and $\Delta \bar{v}^a_{i_1i_2}$.  The first two are needed
for the construction of the compensation charges and the overlap
operator, and the rest for determining the Hamiltonian, and for
evaluating the total energy.

\section{Non-standard Quantities}
The preceding sections have described the details of making a
standard DFT scheme work within the PAW formalism. This section will
focus on what the PAW transform does to quantities needed for
post-processing or expansions to DFT.

It is a big advantage of the PAW method, that it is formally exactly
equivalent to all-electron methods (with a frozen core) but is
computationally comparable to doing pseudopotential calculations. In
pseudopotential approaches, projecting out the core region is handled
by a static projection kernel, while in PAW this projection kernel is
dynamically updated during the SCF-cycle via an expansion of the core
region in a local atomic basis set. This has the drawback for the end
user, the equations for all quantities most be modified to account the
dual basis set description.

\subsection{The External Potential in PAW}
As an example of the principle in accommodating expressions to the PAW
formalism, we will here consider the application of an external
potential in DFT.

Without the PAW transformation, this addition is trivial, as the
desired potential, $v_\ext(\br)$, should simply be added to the
effective KS potential, and the total energy adjusted by the energy
associated with the external potential $E_\ext = \int d\br n(\br)
v_\ext(\br)$.

In PAW, the density decomposes into pseudo and atomic parts, so that
\begin{equation*}
  E_\ext = \int d\br \s{n}(\br)v_\ext(\br) + \sum_a \int d\br \left[n^a(\br) - \s{n}^a(\br)\right]v_\ext(\br).
\end{equation*}
Implying that both a pseudo energy contribution $\s{E}_\ext = \int
d\br \s{n}(\br)v_\ext(\br)$ and atomic corrections $\Delta E_\ext^a =
\int d\br \left[n^a(\br) - \s{n}^a(\br)\right]v_\ext(\br)$ should be
added to the total energy.
\par In PAW, the Hamiltonian has the structure:
\begin{equation*}
  H = \frac{1}{f_n\ket{\psi_n}}\frac{\partial E}{\partial \bra{\psi_n}} = \s{H} + \sum_a \sum_{i_1i_2}\ket{\s{p}^a_{i_1}}\Delta H^a_{i_1i_2} \bra{\s{p}^a_{i_2}}
\end{equation*}
In our case, the extra contributions due to the external potential are:
\begin{equation*}
  \s{H}_\ext(\br) = v_\ext(\br)
\end{equation*}
and
\begin{equation}\label{eq: atomic hamiltonian}
  \Delta H^{a,\ext}_{i_1i_2} = \int d\br v_\ext(\br) \left\{\phi_{i1}^a(\br)\phi_{i2}^a(\br) - \s{\phi}_{i1}^a(\br)\s{\phi}_{i2}^a(\br)\right\}
\end{equation}
Thus we can write the atomic energy contribution as:
\begin{equation*}
  \begin{split}
    \Delta E^a_\ext &= \int d\br v_\ext(\br)\left[n_c^a(\br)-\s{n}_c^a(\br) + \sum_{i_1i_2}D^a_{i_1i_2}\left\{\phi_{i1}^a(\br)\phi_{i2}^a(\br) - \s{\phi}_{i1}^a(\br)\s{\phi}_{i2}^a(\br)\right\}\right]\\
    &= \int d\br v_\ext(\br)\left[n_c^a(\br)-\s{n}_c^a(\br)\right] + \sum_{i_1i_2}D^a_{i_1i_2}\Delta H^{a,\ext}_{i_1i_2}
  \end{split}
\end{equation*}
To evaluate the first term in the last line, the external potential
should be expanded in some radial function at each nuclei e.g. the
gaussians $\tilde{g}^a_{L}$, as the integral of these with the core
densities is already precalculated.

For example, a zero-order (monopole) expansion, equivalent to the
assumption
\begin{equation*}
  v_\ext(\br) \approx v_\ext(\bR^a) \text{ , for } |\br-\bR^a| < r_c^a
\end{equation*}
Leads to the expression:
\begin{equation*}
  \begin{split}
    \Delta E^a_\ext &= v_\ext(\bR^a) (\sqrt{4\pi} Q_{00}^a + \mathcal{Z}^a)\\
    \Delta H^{a,\ext}_{i_1i_2} &= v_\ext(\bR^a) \sqrt{4\pi} \Delta_{00, i_1i_2}^a
  \end{split}
\end{equation*}
Linear external potentials (corresponding to a homogeneous applied
electric field) can be handled exactly by doing an expansion to first
order. This has been used in \gpaw{} in e.g. the paper \cite{Yin2009}.

\subsection{All-electron Density}
During the self-consistency cycle of DFT, the all-electron quantities
are at all times available in principle. In practise, they are never
handled directly, but rather in the composite basis representation of
a global pseudo description augmented by local atomic basis functions.
For some post processing purposes it can however be desirable to
reconstruct all-electron quantities on a single regular grid.

As an example, consider the all-electron density, which can formally
be reconstructed by
\begin{equation*}
  n(\br) = \tilde{n}(\br) + \sum_a \left[ n_c^a(\br) + \sum_{i_1i_2} D_{i_1i_2}^a \left( \phi_{i_1}^a(\br)\phi_{i_2}^a(\br) - \s{\phi}_{i_1}^a(\br)\s{\phi}_{i_2}^a(\br) \right)\right].
\end{equation*}

Transferring this to a uniform grid will of coarse re-introduce the
problem of describing sharply peaked atomic orbitals on a uniform
grid, but as it is only needed for post processing, and not in the
self-consistency, it can be afforded to interpolating the pseudo
density to an extra fine grid, before adding the summed atomic
corrections from the radial grid.

One common use of the all-electron density is for the application of
Bader analysis\cite{Tang2009}. The advantage of applying this to the
all-electron density instead of the pseudo density, is that it can be
proved that the total electron density only has maxima's at the nuclei,
such that there will only be one Bader volume associated with each
atom. This does not hold for the pseudo density, which can result in
detached Bader volumes. In addition, the dividing surfaces found if applied to the pseudo density will be wrong if these intersect the augmentation sphere.

In practice, the reconstructed total density will not integrate
correctly due to the inaccurate description of a uniform grid in the
core regions of especially heavy elements. But as the numerically
exact value of the integral over the atomic corrections are known from
the radial grid description ($=4\pi\sum_{ij}D^a_{ij} \Delta^a_{L, ij}$
), this deficiency can easily be remedied. As long as the density is
correctly described at the dividing surfaces, these will still be
determined correctly.

\subsection{Wannier Orbitals}
When constructing Wannier functions, the only quantities that needs to
be supplied by the DFT calculator are the integrals
$Z_{n_1n_2}^{\mathbf{G}} = \bra{\psi_{n_1}} e^{-\mathbf{G}\cdot \br}
\ket{\psi_{n_2}}$, where $\mathbf{G}$ is one of at most 6 possible (3
in an orthorhombic cell) vectors connecting nearest neighbor cells in
the reciprocal lattice.\cite{Thygesen2005,Ferretti2007}

When introducing the PAW transformation, this quantity can be
expressed as
\begin{equation*}
  Z_{n_1n_2}^{\mathbf{G}} = \bra{\s{\psi}_{n_1}} e^{-\mathbf{G}\cdot \br} \ket{\s{\psi}_{n_2}} + \sum_a \sum_{i_1i_2} P^{a*}_{n_1i_1} P^{a}_{n_2i_2} \left( \bra{\phi^a_{i_1}}e^{-\mathbf{G}\cdot \br}\ket{\phi^a_{i_2}} - \bra{\s{\phi}^a_{i_1}}e^{-\mathbf{G}\cdot \br}\ket{\s{\phi}^a_{i_2}} \right).
\end{equation*}
Even for small systems, the phase of the exponential of the last
integral does not vary significantly over the region of space, where
$\s{p}^a_i$ is non-zero. The integral in the last term can therefore
safely be approximated by
\begin{equation*}
  e^{-\mathbf{G}\cdot \bR^a} \sum_{i_1i_2} P^{a*}_{n_1i_1} P^{a}_{n_2i_2} \sqrt{4\pi}\Delta^a_{00,i_1i_2}.
\end{equation*}

\subsection{Local Properties}
This section describes quantities that can somehow be related to a
specific atom. As the PAW transform utilizes an inherent partitioning
of space into atomic regions, such quantities are usually readily
extractable from already determined atomic attributes, such as the
density matrices or the projector overlaps $P^a_{ni}$, which are by
construction simultaneous expansion coefficients of both the pseudo and the
all-electron wave functions inside the augmentation spheres.

\subsubsection{Projected Density of States}
The projection of the all electron wave functions onto the all
electron partial waves, (i.e. the all electron wave functions of the
isolated atoms) $\phi_i^a$, is within the PAW formalism given by
\begin{equation}
  \langle \phi^a_i | \psi_n\rangle = \langle \tilde \phi^a_i | \tilde \psi_n \rangle + \sum_{a'} \sum_{i_1i_2} \langle \tilde \phi^a_i | \tilde p_{i_1}^{a'} \rangle \Big(\langle \phi_{i_1}^{a'} | \phi_{i_2}^{a'} \rangle - \langle \tilde \phi_{i_1}^{a'} | \tilde  \phi_{i_2}^{a'}\rangle \Big)\langle \tilde p^{a'}_{i_2} | \tilde  \psi_n \rangle
\end{equation}
Using that projectors and pseudo partial waves form a complete basis
within the augmentation spheres, this can be re-expressed as
\begin{equation}
  \langle \phi^a_i | \psi_n \rangle = P^a_{ni} + \sum_{a' \neq a} \sum_{i_1i_2} \langle \tilde \phi^a_i | \tilde p^{a'}_{i_1} \rangle \Delta S^{a'}_{i_1i_2} P^{a'}_{ni_2}
\end{equation}
if the chosen orbital index `i` correspond to a bound state, the
overlaps $\langle \tilde \phi^a_i | \tilde p^{a'}_{i_1} \rangle$,
$a'\neq a$ will be small, and we see that we can approximate
\begin{equation}
  \langle \phi^a_i | \psi_n \rangle \approx \langle \tilde p_i^a | \tilde \psi_n \rangle
\end{equation}
The coefficients $P_{ni}^a = \braket{\s{p}_i^a}{\s{\psi}_n}$, can thus
be used as a qualitative measure of the local character of the true
all electron wave functions. As the coefficients are already
calculated and used in the SCF cycle, it involves no extra
computational cost to determine quantities related directly to these.

These can be used to define an atomic orbital projected density of states
\begin{equation}
  n_i(\varepsilon) = \sum_n \delta(\varepsilon - \epsilon_n) \left|P^a_{ni}\right|^2.
\end{equation}

\subsubsection{Local Magnetic Moments}
As the projection coefficients are simultaneous expansion coefficients
of the pseudo and the all-electron wave functions inside the
augmentation spheres, it can be seen that inside these, the
all-electron density is given by (for a complete set of partial waves)
\begin{equation}
  \label{eq:atom_density}
  n(\br) = \sum_{i_1i_2} D_{i_1i_2}^a \phi_{i_1}^a(\br)\phi_{i_2}^a(\br) + n_c^a(\br)~, \quad |\br - \bR^a| < r_c^a.
\end{equation}

This can be used to assign a local magnetic moment to each atom according to
\begin{equation*}
  M^a = \sum_{i_1i_2} \Delta N^a_{i_1i_2} \left[ D^a_{i_1i_2}(\uparrow) - D^a_{i_1i_2}(\downarrow) \right],
\end{equation*}
where $\Delta N$ is an integration over products of AE waves truncated
to the interior of the augmentation sphere
\begin{equation*}
  \Delta N^a_{i_1i_2} = \int_{\br \in \Omega^a} d\br \phi_{i_1}^a(\br)\phi_{i_2}^a(\br).
\end{equation*}

Note that this will not add up to the total magnetic moment $\int d\br
(n_\uparrow(\br) - n_\downarrow(\br))$, due to the interstitial space
between augmentation spheres, and must be scaled if this is desired.

\subsubsection{LDA + U}
The atom projected density matrix $D^a_{i_1i_2}$ can also be used to
do LDA + U calculations. The \gpaw{} implementation follows the LDA + U
implementation in VASP\cite{Rohrbach2004}, which is based on the
particular branch of LDA + U suggested by Dudarev \emph{et
  al.}\cite{Dudarev1998}, where you set the effective (U-J) parameter.
The key notion is that from \eqref{eq:atom_density} one can define an
(valence-) orbital density matrix
\begin{equation*}
  \hat{\rho}_{i_1i_2}^a = \ket{\phi_{i_1}^a} D^a_{i_1i_2} \bra{\phi_{i_2}^a}.
\end{equation*}
Thus doing LDA + U is a simple matter of picking out the d-type
elements of $D^a$, and adding to the total energy the contribution
\begin{equation}
  \label{eq:hubbard energy}
  \sum_a \sum_{i_1i_2}^\text{d type}\frac{U}{2} \tr \left(D^a_{i_1i_2} - \sum_{i_3} D^a_{i_1i_3} D^a_{i_3i_2} \right)
\end{equation}
and adding the gradient of this to the Hamiltonian
\begin{equation}
  \label{eq:hubbard hamiltonian}
  \sum_a \sum_{i_1i_2}^\text{d type}\ket{\s{p}_{i_1}^a} \frac{U}{2} \left( \delta_{i_1i_2} - 2 D^a_{i_1i_2} \right)\bra{\s{p}_{i_1}^a}
\end{equation}

\subsection{Coulomb Integrals}
When trying to describe electron interactions beyond the level of
standard (semi-) local density approximations, one will often need
Coulomb matrix elements of the type
\begin{equation}
  \label{eq:coulomb-matrix}
  K_{nn',mm'} = (n_{nn'} | n_{mm'}) := \iint \frac{d\br d\br'}{\rr} n^*_{nn'}(\br)n_{mm'}(\br'),
\end{equation}
where the orbital pair density $n_{nn'}(\br) =
\psi_n^*(\br)\psi_{n'}(\br)$.

Such elements are needed in some formulations of vdW functionals
(although not the one implemented in \gpaw), in linear-response TDDFT
(see e.g. \cite{Walter2008}) where only pair densities corresponding
to electron-hole pairs are needed, in exact exchange or hybrid
functionals (see next section) where only elements of the form
$K_{nn',nn'}$ where both indices correspond to occupied states, are
needed, and for GW calculations (see e.g.  \cite{Rostgaard2009}),
where all elements are needed.

Introducing the PAW transformation in \eqref{eq:coulomb-matrix}, the
pair densities partition according to
\begin{equation}
  \label{eq:pair density}
  n_{nn'}(\br) = \s{n}_{nn'}(\br) + \sum_a \left( n^a_{nn'}(\br) - \s{n}^a_{nn'}(\br) \right)
\end{equation}
with the obvious definitions
\begin{align}
  \s{n}_{nn'} =& \s{\psi}_n^*\s{\psi}_{n'} & n^a_{nn'} = & \sum_{i_1i_2} P^{a*}_{ni_1} P^{a}_{n'i_2} \phi^{a*}_{i_1} \phi^{a*}_{i_2} & \s{n}^a_{nn'} = & \sum_{i_1i_2} P^{a*}_{ni_1} P^{a}_{n'i_2} \s{\phi}^{a*}_{i_1} \s{\phi}^{a*}_{i_2}.
\end{align}
Exactly like with the Hartree potential, direct insertion of this in
\eqref{eq:coulomb-matrix} would, due to the non-local nature of the
Coulomb kernel, lead to undesired cross terms between different
augmentation spheres. As before, such terms can be avoided by
introducing some compensation charges, $\s{Z}^a_{nn'}$, chosen such that the
potential of $n^a_{nn'} - \s{n}^a_{nn'} - \s{Z}^a_{nn'}$ are zero
outside their respective augmentation spheres. This is achieved by
doing a multipole expansion and requiring the expansion coefficients
to be zero, and entails a compensation of the form
\begin{equation}
  \label{eq:pair compensation}
  \s{Z}^a_{nn'}(\br) = \sum_L Q^a_{L,nn'} \s{g}^a_L(\br), ~~ Q^a_{L,nn'} = \sum_{i_1i_2} \Delta^a_{L,i_1i_2} P^{a*}_{ni_1} P^{a}_{n'i_2}
\end{equation}
(the constants $\Delta^a_{L,i_1i_2}$ are identical to those in
\eqref{eq: Delta_Lij}).

Introduction of such compensation charges makes it possible to
obtain the clean partitioning
\begin{equation}
  \label{eq:coulomb-matrix-partition}
  K_{nn',mm'} = (\tilde{\rho}_{nn'} | \tilde{\rho}_{mm'}) + 2\sum_a \sum_{i_1i_2i_3i_4} P^{a}_{mi_1} P^{a*}_{ni_2} \Delta C^a_{i_1i_2i_3i_4} P^{a*}_{n'i_3} P^{a}_{m'i_4}.
\end{equation}
Here the last term is a trivial functional of the expansion
coefficients $P^a_{ni}$ involving only the constants $\Delta
C^a_{i_1i_2i_3i_4}$ already precalculated for the atomic corrections
to the Coulomb energy \eqref{eq:coulom tensor 2}. The only
computationally demanding term relates to the Coulomb matrix element
of the smooth compensated pair densities $\tilde{\rho}_{ij} =
\tilde{n}_{ij} + \sum_a \tilde{Z}^a_{ij}$, which are expressible on
coarse grids.

The formally exact partitioning \eqref{eq:coulomb-matrix-partition}
makes it possible, at moderate computational effort, to obtain Coulomb
matrix elements in a representation approaching the infinite basis set
limit. In standard implementations, such elements are usually only
available in atomic basis sets, where the convergence of the basis is
problematic. At the same time, all information on the nodal structure
of the all-electron wave functions in the core region is retained,
which is important due the non-local probing of the Coulomb operator.
In standard pseudopotential schemes, this information is lost, leading
to an uncontrolled approximation to $K_{nn',mm'}$.

As a technical issue, we note that integration over the the Coulomb
kernel $1/\rr$ is done by solving the associated Poisson equation, as
for the Hartree potential, whereby the calculation of each element can
be efficiently parallelized using domain decomposition. The integral
$\int d\br\tilde{\rho}_{nn'}(\br) = \delta_{nn'}$ shows that the
compensated pair densities $\tilde{\rho}_{nn}$ have a non-zero total
charge, which is problematic for the determination of the associated
potential. For periodic systems, charge neutrality is enforced by
subtracting a homogeneous background charge, and the error so
introduced is removed to leading order ($V^{-1/3}$ where $V$ the the
volume of the simulation box) by adding the potential of a missing
probe charge in an otherwise periodically repeated array of probe
charges embedded in a compensating homogeneous background charge. This
can be determined using the standard Ewald technique, and corresponds
to a rigid shift of the potential. For non-periodic systems, all
charge is localized in the box, and the Poisson equation can be solved
by adjusting the boundary values according to a multipole expansion of
the pair density with respect to the center of the simulation box. A
monopole correction is correct to the same order as the correction for
periodic cells, but the prefactor on the error is much smaller, and
leads to converged potentials even for small cells.

\subsubsection{Exact Exchange}
The EXX energy functional is given by
\begin{equation}
  E_\text{xx} =  = -\frac{1}{2} \sum_{nm} f_{n} f_{m}\delta_{\sigma_n,\sigma_m}K_{nm,nm}.
\end{equation}
Terms where $n$ and $m$ both refer to valence states transform in
PAW as in equation \eqref{eq:coulomb-matrix-partition}. Terms where
either index refers to a core orbital can be reduce to
trivial functionals of $P^a_{ni}$, resulting in (see e.g. \cite{Paier2005})
\begin{equation}
  \begin{split}
    E_\text{xx} = -\frac{1}{2}\sum_{nm}^\text{val} &f_{n}f_{m}\delta_{\sigma_n, \sigma_{m}} (\tilde{\rho}_{nm} | \tilde{\rho}_{nm}) \\
    &- \sum_a\left[\sum_\sigma \sum_{i_1i_2i_3i_4} D_{i_1i_3}^{a*}(\sigma) \Delta C_{i_1i_2i_3i_4}^a D_{i_2i_4}^{a}(\sigma) +\sum_{i_1 i_2} D_{i_1 i_2}^a X_{i_1 i_2}^a + E_\text{xx}^{a,\text{c-c}}\right].
  \end{split}
\end{equation}
The term involving the $\Delta C^a$ tensor is the PAW correction for the
valence-valence interaction, and is similar to the correction in the
equivalent expression for the Hartree energy, except that the order of
the indices on the density matrices are interchanged. The term
involving the $X^a$ tensor represents the valence-core exchange
interaction energy. $E_\text{xx}^{a,\text{c-c}}$ is simply the
(constant) exchange energy of the core electrons.

The system independent Hermitian tensor $X_{i_1i_2}^a$ is given by:
\begin{equation}
  \begin{split}
    X_{i_1i_2}^a &= \frac{1}{2}\sum_\alpha^\text{core} \iint d\br d\br'\frac{\phi_{i_1}^{a}(\br)\phi_\alpha^{a,\text{core}}(\br) \phi_{i_2}^{a}(\br')\phi_\alpha^{a,\text{core}}(\br')}{\rr}\\
    &= \sum^\text{core}_{j_c} \sum_l\frac{4\pi}{2l+1}\left(\sum_{m m_c} G^L_{L_1 L_c} G^{L}_{L_2 L_c}\right)\iint drdr'  \frac{r_<^l}{r_>^{l+1}} u^a_{j_1}(r)u^a_{j_c}(r) u^a_{j_c}(r')u^a_{j_2}(r').
\end{split}
\end{equation}

Although the valence-core interaction is computationally trivial to
include, it is not unimportant, giving rise to shifts in the valence
eigenvalues of up to 1eV (though only a few kcal/mol in atomization
energies), and we note that this contribution is unavailable in
pseudopotential schemes. The core-core exchange is simply a reference
energy, and will not affect self-consistency or energy differences.

For the iterative minimization schemes used in real-space and plane
wave codes, the explicit form of the non-local Fock operator
$v^\text{NL}(\br, \br')$ is never needed, and would indeed be
impossible to represent on any realistic grid. Instead only the action
of the operator on a state is needed. As with the Hamiltonian
operator, the action on the pseudo waves is derived via the relation
$f_n \hat{v}^\text{NL} \ket{\tilde{\psi}_n} = \partial E_\text{xx} / \partial \bra{\tilde{\psi}_n}$.
 Referring to \cite{Paier2005} for a derivation, we merely state the result
\begin{multline}\label{eq: nonlocal exchange}
  \h{v}^\text{NL} \ket{\s{\psi}_n} = \sum_m f_m \s{v}_{nm}(\br) \ket{\s{\psi}_m} \\
  + \sum_a \sum_{i_1i_2} \ket{\s{p}_{i_1}^a} \left[ \sum_m v_{nm,i_1i_2}^a P^a_{mi_2} - X^a_{i_1i_2} P^a_{ni_2} - 2 \left( \sum_{i_3i_4}C^a_{i_1i_3i_2i_4}D^a_{i_3i_4} \right) P^a_{ni_2} \right]
\end{multline}
where $\tilde{v}_{nm}$ is the solution of $\nabla^2
\tilde{v}_{nm}(\br) = -4\pi \tilde{\rho}_{nm}(\br)$, and
$v_{nm,i_1i_2}^a = \sum_L \Delta^a_{Li_1i_2}\int d\br\s{g}^a_L(\br)
\tilde{v}_{nm}(\br)$.

Again the computationally demanding first term is related to smooth
pseudo quantities only, which can be accurately represented on coarse
grids, making it possible to do basis set converged self-consistent
EXX calculations at a relatively modest cost. Applying the Fock
operator is however still expensive, as a Poisson equation must be
solved for all pairs of orbitals.

As a technical consideration, note that the effect of the atomic
corrections due to valence-valence, valence-core, and core-core
exchange interactions can simply be incorporated into the standard
equations by redefining equations \eqref{eq:coulom tensor 2},
\eqref{eq:coulom tensor 1}, and \eqref{eq:coulom tensor 0}
respectively, which will also take care of the last two terms in the
Fock operator above. The introduction of the pair orbital compensation
charges does however lead to a non-trivial correction to the Fock
operato; the term proportional to $v^a_{nm, i_1i_2}$. This term also
leads to a distinct contribution when calculating the kinetic energy
via the eigenvalues as done in equation \eqref{eq:kinetic-energy}. The
additional term (besides those related to redefining \eqref{eq:coulom
  tensor 0}--\eqref{eq:coulom tensor 2})
\begin{equation}
  \sum_{nm}f_n\left[f_{m} \delta_{\sigma_n,\sigma_{m}} \int d\br \tilde{v}_{nm}(\br) \psit_n^*(\br)\psit_{m}(\br) - \sum_a\sum_{i_1i_2} P_{ni_1}^aP_{mi_2}^a v^a_{nm,i_1i_2}\right],
\end{equation}
should be added to the right hand side of \eqref{eq:kinetic-energy} on
inclusion of exact exchange.

In a similar fashion, the compensation charges leads to an additional
force contribution in equation \eqref{eq: force no exx} given by
\begin{equation}\label{eq:paw-exx-force}
  \begin{split}
    \mathbf{F}^a_{xx} = \sum_{nm} f_{n}f_{nm}\delta_{\sigma_{n} \sigma_{m}}\Bigg\{ &\int d\br' \tilde{v}_{nm}(\br') \sum_{i_1i_2} P^{a*}_{ni_1}P^{a}_{mi_2}\sum_L\Delta_{Li_1i_2}\frac{\partial \tilde{g}^a_L(\br')}{\partial \bR^a}\\
    & + \sum_{i_1i_2} v^a_{n_1n_2i_1i_2}\left( P^{a*}_{n
        i_1}\braket{\frac{d \pt^a_{i_2}}{d\bR^a}}{\psit_{m}} +
      \braket{\psit_{n}}{\frac{d \pt^a_{i_1}}{d\bR^a}} P^a_{m i_2}
    \right)\Bigg\}.
  \end{split}
\end{equation}

\subsubsection{Optimized Effective Potential}
The optimized effective potential (OEP) method, is a way of converting
the non-local Fock operator $\hat{v}^\text{NL}_x$ into a local form
$\hat{v}^\text{L}_x = v^\text{L}_x(\br)$.

One way to derive the OEP equations in standard KS-DFT, is to use
perturbation theory along the adiabatic connection (G\"orling-Levy
perturbation theory \cite{Gorling1994}).

On converting the OEP equation to the PAW formalism, it should be
remembered that local potentials in PAW transform to a local pseudo
part plus non-local atomic corrections. Hence we want to arrive at a
potential of the form
\begin{equation}\label{eq: local exchange}
  \h{v}_x^\text{L} = \s{v}_x^\text{L}(\br) + \sum_a \sum_{i_1i_2} \ket{\s{p}_{i_1}^a}\Delta v_{i_1i_2}^a \bra{\s{p}_{i_2}^a},
\end{equation}
where both the pseudo part $\s{v}_x^\text{L}$ as well as the
coefficients $\Delta v_{i_1i_2}^a$ should be determined.

The derivation is more or less straight forward, if one remembers the
the PAW KS equation is a generalized eigenvalue problem, that the
variational quantity is the pseudo orbitals, and that the first order
shift in the density has both a pseudo and an atomic part. The result is
\begin{subequations}\label{eq: paw oep2}
\begin{align}
  \sum_n f_n \s{\psi}_n^*(\br) \sum_{m\neq n} \s{\psi}_m(\br) \frac{\bra{\s{\psi}_m} \h{v}_x^\text{NL} - \h{v}_x^\text{L} \ket{\s{\psi}_n}}{\epsilon_n - \epsilon_m}  + c.c. &= 0\\
  \sum_n f_n P_{ni_1}^{a*}\sum_{m\neq n}P_{mi_2}^{a}
  \frac{\bra{\s{\psi}_m} \h{v}_x^\text{NL} - \h{v}_x^\text{L}
    \ket{\s{\psi}_n}}{\epsilon_n - \epsilon_m} + c.c. &= 0
\end{align}
\end{subequations}
where $\hat{v}^\text{NL}_x$ is the non-local exchange operator of
equation \eqref{eq: nonlocal exchange} and $\hat{v}^\text{L}_x$ is the
local version in \eqref{eq: local exchange}.

These can be solved iteratively starting from a local density-function
approximation to the exchange potential in the spirit of
\cite{Kummel2003}.

It might seem that OEP is just extra work on top of the already
expensive non-local operator, but it can in some cases be faster, as
the number of SCF iterations in the KS cycle are greatly reduced.

\clearpage
\addcontentsline{toc}{section}{References}
\bibliographystyle{unsrt}

\end{document}